\documentclass[11pt]{article}
\usepackage{typearea}\typearea{12}
\usepackage{epsfig,amsmath,amsfonts,amssymb}
\usepackage[dvipdfmx]{hyperref}
\hypersetup{hidelinks}
\newtheorem{theorem}{Theorem}[section]
\newtheorem{lemma}[theorem]{Lemma}

\newtheorem{proposition}[theorem]{Proposition}
\newtheorem{definition}[theorem]{Definition}

\makeatletter
\@addtoreset{equation}{section}
\makeatother

\makeatletter
\long\def\@makecaption#1#2{{\small
\advance\leftskip1cm
\advance\rightskip1cm
\vskip\abovecaptionskip
\sbox\@tempboxa{#1: #2}%
\ifdim \wd\@tempboxa >\hsize
 #1: #2\par
\else
\global \@minipagefalse
\hb@xt@\hsize{\hfil\box\@tempboxa\hfil}%
\fi
\vskip\belowcaptionskip}}
\makeatother
\def\eq#1\en{\begin{equation}#1\end{equation}}  
\def\eqa#1\ena{\begin{align}#1\end{align}}
\def\eqg#1\eng{\begin{gather}#1\end{gather}}
\newcommand{\lb}[1]{\label{e:#1}}
\newcommand{\rlb}[1]{\eqref{e:#1}} 
\newcommand{\nl}{\notag\\}

\newcommand{\snorm}[1]{\Vert#1\Vert}

\newcommand{\sbkt}[1]{\langle#1\rangle}

\newcommand{\sumtwo}[2]%
{\mathop{\sum_{#1}}_{#2}}
\newcommand{\sumthree}[3]%
{\mathop{\mathop{\sum_{#1}}_{#2}}_{#3}}
\newcommand{\sumfour}[4]%
{\mathop{\mathop{\mathop{\sum_{#1}}_{#2}}_{#3}}_{#4}} 
\newcommand{\prodtwo}[2]%
{\mathop{\prod_{#1}}_{#2}}
\newcommand{\mintwo}[2]%
{\mathop{\min_{#1}}_{#2}}
\newcommand{\maxtwo}[2]%
{\mathop{\max_{#1}}_{#2}}
\newcommand{\maxthree}[3]%
{\mathop{\mathop{\max_{#1}}_{#2}}_{#3}}
\newcommand{\limtwo}[2]%
{\mathop{\lim_{#1}}_{#2}}
\newcommand{\suptwo}[2]%
{\mathop{\sup_{#1}}_{#2}}
\newcommand{\supthree}[3]%
{\mathop{\mathop{\sup_{#1}}_{#2}}_{#3}}
\newcommand{\supfour}[4]%
{\mathop{\mathop{\mathop{\sup_{#1}}_{#2}}_{#3}}_{#4}} 
\newcommand{\inftwo}[2]%
{\mathop{\inf_{#1}}_{#2}}
\newcommand{\infthree}[3]%
{\mathop{\mathop{\inf_{#1}}_{#2}}_{#3}}
\newcommand{\inffour}[4]%
{\mathop{\mathop{\mathop{\inf_{#1}}_{#2}}_{#3}}_{#4}} 
\newcommand\calA{{\cal A}}

\newcommand\calH{{\cal H}}

\newcommand\calP{{\cal P}}

\newcommand\calS{{\cal S}}

\newcommand{\sfJ}{\mathsf{J}}

\newcommand{\bbC}{\mathbb{C}}

\newcommand{\bbN}{\mathbb{N}}

\newcommand{\bbR}{\mathbb{R}}
\newcommand{\bbZ}{\mathbb{Z}}

\newcommand{\up}{\uparrow}

\newcommand{\qedm}{\rule{1.5mm}{3mm}}

\newcommand{\pmat}[1]{\begin{pmatrix}#1\end{pmatrix}}
\newcommand{\ket}[1]{|#1\rangle}

\newcommand{\La}{\Lambda}

\newcommand{\hQ}{\hat{Q}}

\newcommand{\hA}{\hat{A}}
\newcommand{\hB}{\hat{B}}
\newcommand{\hC}{\hat{C}}
\newcommand{\hH}{\hat{H}}
\newcommand{\hI}{\hat{I}}

\newcommand{\hW}{\hat{W}}
\newcommand{\hX}{\hat{X}}
\newcommand{\hY}{\hat{Y}}
\newcommand{\hZ}{\hat{Z}}

\newcommand{\hS}{\hat{S}}

\newcommand{\hh}{\hat{h}}

\newcommand{\Tr}{\operatorname{Tr}}

\newcommand{\bA}{\boldsymbol{\hat{A}}}
\newcommand{\bB}{\boldsymbol{\hat{B}}}
\newcommand{\bC}{\boldsymbol{\hat{C}}}
\newcommand{\bD}{\boldsymbol{\hat{D}}}
\newcommand{\bE}{\boldsymbol{\hat{E}}}
\newcommand{\bF}{\boldsymbol{\hat{F}}}
\newcommand{\bh}{\boldsymbol{\hat{h}}}

\newcommand{\la}{\lambda}

\newcommand{\Zd}{\mathbb{Z}^d}

\newcommand{\ri}{\mathrm{i}}
\newcommand{\hSt}{\hat{S}_{\rm tot}}

\newcommand{\bea}{\boldsymbol{e}_\alpha}

\newcommand{\wid}{\operatorname{Wid}}
\newcommand{\wido}{\operatorname{Wid}_1}
\newcommand{\supp}{\operatorname{Supp}}

\newcommand{\jodd}{\text{$j$ odd}}
\newcommand{\jeven}{\text{$j$ even}}

\newcommand{\qA}{q_{\bA}}
\newcommand{\qB}{q_{\bB}}
\newcommand{\cB}{c_{\bB}}

\newcommand{\PL}{\calP_\La}
\newcommand{\PZd}{\calP_{\Zd}}

\newcommand{\Jx}{J_\mathrm{X}}
\newcommand{\Jy}{J_\mathrm{Y}}
\newcommand{\Jz}{J_\mathrm{Z}}
\newcommand{\Jxx}{J_\mathrm{XX}}
\newcommand{\Jyy}{J_\mathrm{YY}}
\newcommand{\Jzz}{J_\mathrm{ZZ}}
\newcommand{\Jxy}{J_\mathrm{XY}}
\newcommand{\Jyx}{J_\mathrm{YX}}
\newcommand{\Jyz}{J_\mathrm{YZ}}
\newcommand{\Jzy}{J_\mathrm{ZY}}
\newcommand{\Jzx}{J_\mathrm{ZX}}
\newcommand{\Jxz}{J_\mathrm{XZ}}

\newcommand{\hx}{h_\mathrm{X}}
\newcommand{\hy}{h_\mathrm{Y}}
\newcommand{\hz}{h_\mathrm{Z}}
\newcommand{\km}{k_\mathrm{max}}

\newcommand{\kn}{k}

\newcommand{\AXY}{\calA^{\hX\hX\app\hY}}

\newcommand{\AYX}{\calA^{\hY\hY\app\hX}}
\newcommand{\AYZ}{\calA^{\hY\hY\app\hZ}}

\newcommand{\kk}{_{(\kn-1,1)\app(\kn-1,0)}}

\newcommand{\dive}[1]{\bigl|#1\bigr|}

\newcommand{\hsp}{\hspace{3pt}}
\newcommand{\ZZ}[2]{\biggl(\hsp\bigotimes_{\nu=#1}^{#2}\hZ_{(\nu,0)}\biggr)}

\newcommand{\eo}{\boldsymbol{e}_1}

\newcommand{\tq}{\tilde{q}}
\newcommand{\newQ}{\mathcal{Q}}

\newcommand{\app}{\leadsto}

\usepackage{color}
\definecolor{fluorescentpink}{rgb}{1.0, 0.08, 0.58}
\definecolor{forestgreen}{rgb}{0.13, 0.55, 0.13}

\begin{document}

\noindent
{\Large\bf 
The $S=\frac{1}{2}$ XY and XYZ models on the two or higher dimensional hypercubic lattice do not possess nontrivial local conserved quantities}

\renewcommand{\thefootnote}{\fnsymbol{footnote}}
\medskip\noindent
Naoto Shiraishi\footnote{Faculty of Arts and Sciences, University of Tokyo, 3-8-1 Komaba, Meguro-ku, Tokyo,
Japan.} and Hal Tasaki\footnote{%
Department of Physics, Gakushuin University, Mejiro, Toshima-ku, 
Tokyo 171-8588, Japan.
}
\renewcommand{\thefootnote}{\arabic{footnote}}
\setcounter{footnote}{0}

\begin{quotation}
\small\noindent
We study the $S=\frac{1}{2}$ quantum spin system on the $d$-dimensional hypercubic lattice with $d\ge2$ with uniform nearest-neighbor interaction of the XY or XYZ type and arbitrary uniform magnetic field.
By extending the method recently developed for quantum spin chains, we prove that the model possesses no local conserved quantities except for the trivial ones, such as the Hamiltonian.
This result strongly suggests that the model is non-integrable.
We note that our result applies to the XX model without a magnetic field, which is one of the easiest solvable models in one dimension.

\medskip\noindent
{\em There are a 23-minute video that discusses the main results of the present work}
\\\url{https://youtu.be/uhj7mPBmMn4}
\\{\em and a 44-minute video that explains the basic ideas of the proof}
\\\url{https://youtu.be/gqHBkxdHDoU}
\end{quotation}

\tableofcontents

\section{Introduction}\label{S:introduction}
Mathematical investigations into quantum many-body systems may be roughly divided into two approaches.
In the first, one focuses on exactly solving specific, concrete models to extract detailed and precise information about the system\footnote{
There are quantum many-body systems whose ground states (and limited excited states) can be obtained exactly.
We do not call such systems exactly solvable in the present discussion.
} \cite{JimboMiwa,Fadeev,Baxter,Takahashi}. 
These exactly solvable models are often referred to as integrable models. 
In the second approach, the goal is to establish rigorous, physically meaningful results that apply more generally to a broad class of models \cite{Simon,Tasaki}. Importantly, these results do not differentiate between models that are integrable and those that are not.

There are, however, physically significant properties, such as quantum chaos and the energy eigenstate thermalization hypothesis (ETH)\footnote{\label{fn:scar}
In short, the ETH asserts that every energy eigenstate of a non-random, non-integrable system is thermal in the sense that it is macroscopically indistinguishable from the thermal equilibrium state.
It should be stressed, however, that the ETH is not ubiquitous in non-integrable systems.
It was noted both experimentally~\cite{Bernienetal2017} and (independently) theoretically~\cite{ShiraishiMori2017,MoriShiraishi2017} that some energy eigenstates in a certain non-integrable system are not thermal while many other energy eigenstates are.
Such exceptional energy eigenstates were later named quantum many-body scar states and have been attracting considerable interest~\cite{TurnerMichailidisAbaninSerbynPapic2018, MoudgalyaRachelBernevigRegnault2018, MoudgalyaRegnaultBernevig2018, LinMotrunich2018, HoChoiPichlerLukin2019, Shiraishi2019A, MarkLinMotrunich2020,MoudgalyaRegnaultBernevig2020,SerbynAbaninPapic2021}.
}, that are conjectured to take place exclusively in non-integrable systems \cite{Deutsch1991,Srednicki1994,RigolSrednicki2012,DAlessioKafriPolkovnikovRigol2016}. 
Despite their importance, mathematically rigorous theories that specifically address non-integrable systems and uncover these complex properties remain scarce and in high demand.
In 2019, one of the present authors (N.S.) developed a new strategy and proved that the one-dimensional $S=\frac{1}{2}$ quantum XYZ model under a magnetic field does not possess any nontrivial local conserved quantities \cite{Shiraishi2019}.
As far as we know, this was the first rigorous demonstration that quantum many-body systems in a concrete class are not solvable or, to be more precise, exhibit a property that is not shared by any of the exactly solved models.\footnote{One may interpret Bouch's result in 2015 as a precursor \cite{Bouch2015}.  See Appendix~\ref{s:ITE}.}
The method was extended to the $S=\frac{1}{2}$ quantum Ising chain under a magnetic field \cite{Chiba2024a}, the PXP model \cite{ParkLee2024a}, and the $S=\frac{1}{2}$ spin chains with next-nearest-neighbor interactions \cite{Shiraishi2024}.
The article \cite{Shiraishi2024} by one of us (N.S.) provides a detailed exposition of the method applied to the prototypical case of the XYZ-h model as well as a new result for chains with next-nearest-neighbor interactions.
More recently, the method was applied to test for the presence/absence of nontrivial local conserved quantities exhaustively in general classes of spin chains, namely, the $S=1$ bilinear/biquadratic chains (with an anisotropy) \cite{ParkLee2024b,HokkyoYamaguchiChiba2024}, the $S=\frac{1}{2}$ chain with nearest neighbor symmetric interactions \cite{YamaguchiChibaShiraishi2024a,YamaguchiChibaShiraishi2024b}, and, finally,  the $S=\frac{1}{2}$ chain with nearest and next-nearest neighbor symmetric interactions \cite{Shiraishi2025}.
The same method has also been useful in providing explicit characterizations of local conserved quantities in integrable models \cite{NozawaFukai2020,YamadaFukai2023,Fukai2023,Fukai2024}.

It may not be fruitful to discuss whether the absence of nontrivial local conserved quantities established in \cite{Shiraishi2019,Chiba2024a,ParkLee2024a,Shiraishi2024,ParkLee2024b,HokkyoYamaguchiChiba2024,YamaguchiChibaShiraishi2024a,YamaguchiChibaShiraishi2024b,Shiraishi2025} ``proves the non-integrability'' of these models.
The answer to such a question depends on how one defines the integrability of a quantum many-body system.
See, e.g., \cite{GrabowskiMathieu1995,CauxMossel2011}.
Interestingly, recent exhaustive investigations of the $S=1$ chain with the standard bilinear and biquadratic interactions \cite{ParkLee2024b,HokkyoYamaguchiChiba2024}, the $S=\frac{1}{2}$ chain with nearest neighbor symmetric interactions \cite{YamaguchiChibaShiraishi2024a,YamaguchiChibaShiraishi2024b}, and the $S=\frac{1}{2}$ chain with nearest and next-nearest neighbor symmetric interactions have revealed that all the models in these classes, except for those were already known to be integrable, do not possess nontrivial local conserved quantities.
There seems to be an empirical rule that any simple translation invariant model is either integrable or lacks nontrivial local conserved quantities.
It is challenging to understand deeper connections between the absence of nontrivial local conserved quantities and other properties that are expected to take place in non-integrable systems.

\medskip
In the present work, we extend the method developed in \cite{Shiraishi2019,Shiraishi2024} to models in higher dimensions.
We study the general $S=\frac{1}{2}$ quantum spin model on the $d$-dimensional hypercubic lattice with $d\ge2$ with the standard translation-invariant Hamiltonian \rlb{H} that has (possibly anisotropic) nearest neighbor exchange interactions and an arbitrary magnetic field.
By only assuming $\Jx\ne0$ and $\Jy\ne0$, we prove that the model has no nontrivial local conserved quantities whose support has a width larger than 2.
This means that even the simplest XX model with the Hamiltoian $\hH_{\rm XX}=-\frac{1}{2}\sum_{u,v\ \text{s.t.}\ |u-v|=1}(\hX_u\hX_v+\hY_u\hY_v)$ (see Section~\ref{S:main} for notation), which is one of the easiest solvable models in one-dimension, has no nontrivial local conserved quantities in dimensions two or higher.
Our result provides strong support to the intuition, shared by most researchers for decades, that quantum models become ``less solvable'' as the dimension increases.
It is quite likely that the only standard model that allows nontrivial local conserved quantities in higher dimensions is the Ising model under the magnetic field in the direction parallel to that of the exchange interaction, which is nothing but the classical Ising model.\footnote{
There are some ``exotic'' (but important) models in higher dimensions that have nontrivial conserved quantities.
Notable examples include the models that have the generalized Briegel-Raussendorf states \cite{BriegelRaussendorf2001,RaussendorfBriegel2001} (the cluster states) as their ground states (see, e.g., \cite{Tasaki}), Kitaev's toric code model \cite{Kitaev2003}, and the Kitaev honeycomb model \cite{Kitaev2006}.
}

Our strategy of the proof is a natural extension of that proposed in the original work \cite{Shiraishi2019}.
We start by writing down the system of linear equations that fully characterizes a local conserved quantity.
We then make use of the procedure introduced in \cite{Shiraishi2019} that we call ``shift'' to reduce the task of characterization to a problem in an essentially one-dimensional setting.
This part is intrinsic to the treatment of models in two or higher dimensions.
We finally analyze, again following the strategy in \cite{Shiraishi2019}, this one-dimensional problem to conclude that there are no nontrivial local conserved quantities.
We note that this final step is considerably simpler compared with the corresponding step in \cite{Shiraishi2019,Chiba2024a,ParkLee2024a,Shiraishi2024,ParkLee2024b,HokkyoYamaguchiChiba2024,YamaguchiChibaShiraishi2024a,YamaguchiChibaShiraishi2024b,Shiraishi2025} for one-dimensional models.
This does not mean we have a better mathematical strategy, but it means ruling out conservation laws is easier in higher dimensions.

When we were preparing the present paper, we learned that Yuuya Chiba had completed a proof of the absence of nontrivial local conserved quantities in the $S=\frac{1}{2}$ Ising model under the transverse magnetic and longitudinal fields in two or higher dimensions \cite{Chiba2024b}.
More recently, after the present work was made public, the absence of local conserved quantities in two or higher dimensions was proved for the quantum compass model by Futami and one of us (H.T.) \cite{FutamiTasaki} and for the Hubbard model by Futami \cite{Futami}.

\medskip

The present paper is self-contained and assumes only a basic background in quantum spin systems.
In particular, we do not assume familiarity with the method applied to one-dimensional models.
We do not employ specialized notation that was developed in \cite{Shiraishi2019,Shiraishi2024} to deal efficiently with complicated mathematical puzzles that one encounters in one-dimensional models.

The present paper is organized as follows.
In Section~\ref{S:main}, we fix notation, define our model, and discuss the main theorems.
Section~\ref{S:proof} is devoted to the proof of the theorems.
After describing the basic strategy in Section~\ref{s:basic}, we prove in Section~\ref{S:reduction} basic lemmas that allow us to reduce the problem to that in one dimension.
The one-dimensional puzzle is then solved in Sections~\ref{S:k=3} to \ref{S:CXX}.
Section~\ref{S:k=2} is devoted to the proof of a theorem for conserved quantities with a small support.
In Section~\ref{S:discussion}, we discuss several possibilities for extending the present approach to uncover the complex nature exhibited by non-integrable quantum many-body systems.
We provide some results for the four among such topics, namely, the absence of quasi-local conserved quantities, the limitation to spectrum generating algebra (SGA), lower bounds for the Lanczos coefficients characterizing operator growth, and the singularity in complex-time evolution of opeartoras in Appendices~\ref{S:QuasiLocal}, \ref{S:SGA}, \ref{S:OG}, and \ref{s:ITE}.

\medskip\noindent
{\em Remark:}\/
After the earlier version of the present paper was made public, Hokkyo \cite{Hokkyo2025} developed an efficient, rigorous scheme for establishing the absence of local conserved quantities in a general class of quantum spin systems (mostly in one dimension).
His scheme drastically simplifies the proof of the absence of nontrivial local conserved quantities in some of the models mentioned above.
In particular, one may prove the main results of the present paper by explicitly justifying the conditions for the general theorem in Section VI of \cite{Hokkyo2025}.
See also \cite{Hokkyo2025B} and \cite{ShiraishiYamaguchi} for more recent progress in general theories that distinguish between integrable and non-integrable quantum systems.

\section{Definitions and the main theorem}\label{S:main}
\begin{figure}
\centerline{\epsfig{file=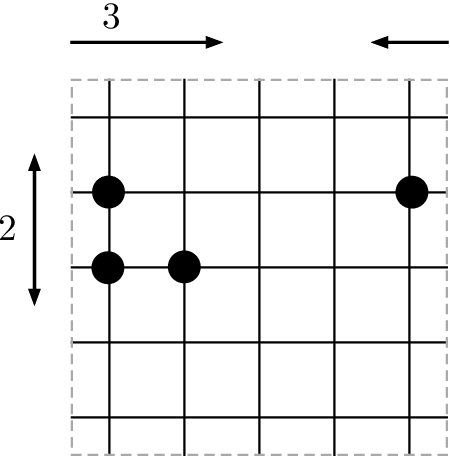,width=3.5truecm}}
\caption[dummy]{
The lattice $\La$ with $d=2$ and $L=5$, namely, the $5\times 5$ square lattice.
Note that the figure represents the whole lattice.
A subset $S$ with four elements is depicted by black disks.
Taking into account the periodic boundary conditions, we see $\wido S=3$ and $\wid_2S=2$.
}
\label{f:widths}
\end{figure}

Let $\La=\{1,\ldots,L\}^d$ be the $d$-dimensional $L\times\cdots\times L$ hypercubic lattice with periodic boundary conditions, where $d\ge2$.
For a nonempty subset $S\subset\La$, we define its width in the $\alpha$-direction (where $\alpha\in\{1,\ldots,d\}$), denoted as $\wid_\alpha S$, as the minimum $k$ such that 
\eq
0\le(u)_\alpha-a\le k-1\ ({\rm mod}\ L),
\lb{0<a-a}
\en
for every $u\in S$ with some $a\in\{1,\ldots,L\}$.
Here $(u)_\alpha$ denotes the $\alpha$-th coordinate of $u$.
See Figure~\ref{f:widths}.
We also define
\eq
\wid S=\max_{\alpha\in\{1,\ldots,d\}}\wid_\alpha S.
\en

We consider the spin system with $S=\frac{1}{2}$ on the lattice $\La$.
Let
\eq
\hX=\pmat{0&1\\1&0},\quad
\hY=\pmat{0&-\ri\\\ri&0},\quad
\hZ=\pmat{1&0\\0&-1},
\en
denote the Pauli matrices for a single spin.
These three matrices and the identity $\hI$ span the space of operators of a single spin.
For $u\in\La$, we denote by $\hX_u$, $\hY_u$, $\hZ_u$, and $\hI_u$ the copies of these matrices at site $u$.

We study the Hamiltonian
\eqa
 \hH=&-\frac{1}{2}\sumtwo{u,v\in\La}{(|u-v|=1)}\bigl\{\Jx\,\hX_u\hX_v+\Jy\,\hY_u\hY_v+\Jz\,\hZ_u\hZ_v\bigr\}
 \nl&-\sum_{u\in\La}\bigl\{\hx\,\hX_u+\hy\,\hY_u+\hz\,\hZ_u\bigr\},
 \lb{H}
 \ena
where $\Jx,\Jy,\Jz\in\bbR$ are the exchange ineraction constants and $(\hx,\hy,\hz)\in\bbR^3$ represents the external magnetic field.
To be rigorous, the term $\hX_u\hX_v$, for example, should be $\hX_u\otimes\hX_v\otimes(\bigotimes_{w\in\La\backslash\{u,v\}}\hI_w)$, but we omit the identity and the tensor product symbol.

In the present work, we assume $\Jx\ne0$ and $\Jy\ne0$.
Our model is the XY model when $\Jz=0$ and the XYZ model when $\Jz\ne0$.
We make no assumptions on the magnetic field $(\hx,\hy,\hz)$.

By a product of Pauli matrices (which we shall often refer to as simply a product), we mean an operator of the form
\eq
\bA=\Bigl(\bigotimes_{u\in S}\hA_u\Bigr)\otimes\Bigl(\bigotimes_{v\in\La\backslash S}\hI_v\Bigr),
\lb{A}
\en
where $S\subset\La$ is nonempty and $\hA_u\in\{\hX_u,\hY_u,\hZ_u\}$.
We denote $S$ as $\supp\bA$ and refer to it as the support of $\bA$.
As above, we shall omit the identity and write \rlb{A} as $\bA=\bigotimes_{u\in S}\hA_u$.
We denote the set of all products of Pauli matrices on $\La$ by $\PL$.
Note that the elements of $\PL$, with the identity $\hI$, span the whole space of operators of the spin system on $\La$.
We define the widths of $\bA$ by
\eq
\wid_\alpha\bA=\wid_\alpha\supp\bA,\quad\wid\bA=\wid\supp\bA,
\en
for $\alpha\in\{1,\ldots,d\}$.

We are almost ready to state our theorem.
Fix a constant $\km$ such that 
\eq
2\le \km\le\frac{L}{2}.
\lb{k}
\en
We write the candidate of a local conserved quantity as
 \eq
 \hQ=\sumtwo{\bA\in\PL}{(\wid\bA \le \km)}\qA\,\bA,
 \lb{Q}
 \en
 where $\qA\in\bbC$ are coefficients.
 We assume there exists at least one $\bA\in\PL$ such that $\qA\ne0$ and $\wid\bA=\km$.
 We do not assume any symmetry, such as the translation symmetry, on the coefficients $\qA$.
 Note that \rlb{Q} may express any operator if there is no condition on $\wid\bA$.
 We are requiring that $\hQ$ is a superposition of products of Pauli matrices with the maximum width $\km$.
 
 We say that $\hQ$ is a local conserved quantity if and only if
 \eq
 [\hH,\hQ]=0.
 \lb{HQ}
 \en
 Then, the following theorem is the main conclusion of the present paper.
 \begin{theorem}\label{T:main}
 There are no local conserved quantities $\hQ$ with $3\le\km\le L/2$.
 \end{theorem}
 Of course, the Hamiltonian $\hH$ is a local conserved quantity with $\km=2$.
 Indeed, we can prove that the two-body part of any local conserved quantity $\hQ$ with $\km=2$ must be a constant multiple of the two-body part of the Hamiltonian.
Note that the one-body part of $\hQ$ is undetermined when there exists a local conserved quantity with $\km=1$, such as the total magnetization.
The following theorem is, therefore, optimal in the general setting.
\begin{theorem}\label{T:k=2}
Any local conserved quantity $\hQ$ with $\km=2$ is written as
\eq
\hQ=\eta\hH+\hQ^{(1)},
\lb{Q=H}
\en
with $\eta\in\bbC\backslash\{0\}$.
Here, $\hQ^{(1)}$ is a one-body operator, i.e., a linear combination of $\hX_u$, $\hY_u$, and $\hZ_u$ with $u\in\La$.
 \end{theorem}
 
 \noindent{\bf Remarks:}
\par\noindent
1.~Note that $\hH^2$ is a conserved quantity with $\km=(L/2)+2$ for even $L$.
Exactly as in the original work \cite{Shiraishi2019}, we see that the restriction $\km\le L/2$ in Theorem~\ref{T:main} is optimal.

\medskip\par\noindent
2.~One can define the notion of local conserved quantities in the spin model on the infinite hypercubic lattice $\Zd$ and prove Theorem~\ref{T:main} (with the condition $3\le\km\le L/2$ replaced by $\km\ge3$) and Theorem~\ref{T:k=2}.
See Appendix~\ref{S:QuasiLocal}, in particlar, footnote~\ref{fn:localforZd}.
  
 \section{Proof}\label{S:proof}
 \subsection{Basic strategy and some notations}\label{s:basic}
We shall closely follow the strategy developed in \cite{Shiraishi2019,Shiraishi2024}.
For a product $\bA\in\PL$, we express its commutator with the Hamiltonian as a linear combination of products as
\eq
[\hH,\bA]=\sum_{\bB\in\PL}\la_{\bA,\bB}\,\bB.
\lb{HAPL}
\en
The coefficients $\la_{\bA,\bB}$ are readily determined by \rlb{H} and the basic commutation relations between the Pauli matrices.
We then note that the commutator $[\hH,\hQ]$ for a general operator of the form \rlb{Q} is expressed as
\eq
[\hH,\hQ]=\sumtwo{\bA\in\PL}{(\wid\bA \le \km)}\qA\,[\hH,\bA]=\sumtwo{\bB\in\PL}{(\wid\bB\,\le\,\km+1)}\cB\,\bB,
\lb{HQexp}
\en
where the coefficients are 
\eq
\cB=\sum_{\bA\in\PL}\la_{\bA,\bB}\,\qA.
\lb{cB2}
\en
See the discussion that follows \rlb{App} below for the upper bound of $\wid\bB$.
Then the condition \rlb{HQ}, which means that $\hQ$ is conserved, is equivalent to
 \eq
\cB=0,
 \lb{c=0}
 \en
 for all $\bB\in\PL$.
 This gives a coupled linear equations for $\qA$ with $\bA\in\PL$.
 The strategy of the proof of Theorem~\ref{T:main} is to use the relations \rlb{c=0} for a suitably chosen set of $\bB$ and deduce that $\qA=0$ for any $\bA\in\PL$ with $\wid\bA=\km$.
 This contradicts the assumption that $\km$ is the maximum width of products that constitute the conserved quantity $\hQ$, and the theorem follows.
 
We need to evaluate commutators $[\hH,\bA]$ for various $\bA\in\PL$.
For the reader's convenience, let us summarize the basic properties of the Pauli matrices:\footnote{
Note to the reader of \cite{Shiraishi2024}:
 In \cite{Shiraishi2024}, the product of Pauli matrices of a single spin is denoted by a dot as $\hX\cdot\hY$ to avoid confusion with the (convenient) shorthand notation $\hX\hY\hZ_3=\hX_1\otimes\hY_2\otimes\hZ_3$.
We do not use these notations in the present paper.
 }
 \eqg
 \hX^2=\hY^2=\hZ^2=\hI,\\
 \hX\hY=-\hY\hX=\ri\hZ,\quad
 \hY\hZ=-\hZ\hY=\ri\hX,\quad
 \hZ\hX=-\hX\hZ=\ri\hY.
 \lb{XYZ}
 \eng
 These, in particular, imply
 \eq
 [\hW,\hW']=2\hW\hW'=2\ri\,\sigma(\hW,\hW')\,\dive{\hW\hW'},
 \lb{WW}
 \en
 for any $\hW,\hW'\in\{\hX,\hY,\hZ\}$ with $\hW\ne\hW'$.
 Here,
 \eqg
 \sigma(\hX,\hY)=\sigma(\hY,\hZ)=\sigma(\hZ,\hX)=1,\nl
 \sigma(\hY,\hX)=\sigma(\hZ,\hY)=\sigma(\hX,\hZ)=-1,
 \lb{sign}
 \eng
 are the standard sign factors, and the divestment operation \cite{Shiraishi2024}, which divests the phase factor of a Pauli matrix, is defined by
 \eq
 \dive{a\hW}=|a|\,\hW,
 \lb{divest}
 \en
 for $\hW\in\{\hX,\hY,\hZ\}$ and $a\in\bbC$.
 
\begin{figure}
\centerline{\epsfig{file=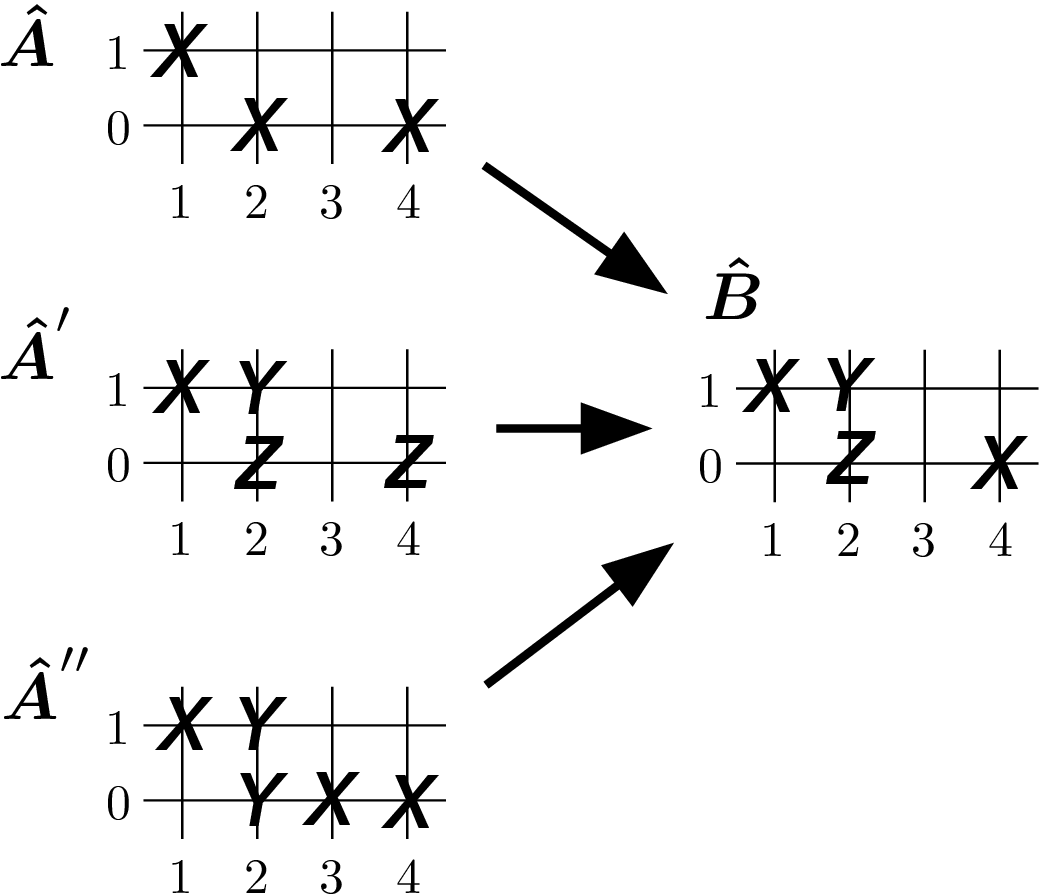,width=7truecm}}
\caption[dummy]{
The figure represents a small portion of the square lattice, where the numbers indicate the first and second coordinates.
The product $\bB=\hX_{(1,1)}\hY_{(2,1)}\hZ_{(2,0)}\hX_{(4,0)}$ is generated, for example, from $\bA=\hX_{(1,1)}\hX_{(2,0)}\hX_{(4,0)}$ with $\hY_{(2,1)}\hY_{(2,0)}$, from $\bA'=\hX_{(1,1)}\hY_{(2,1)}\hZ_{(2,0)}\hZ_{(4,0)}$ with $\hY_{(4,0)}$, and from $\bA''=\hX_{(1,1)}\hY_{(2,1)}\hY_{(2,0)}\hX_{(3,0)}\hX_{(4,0)}$ with $\hX_{(2,0)}\hX_{(3,0)}$.
Note that $\supp\bB\supsetneqq\supp\bA$, $\supp\bB=\supp\bA'$, and $\supp\bB\subsetneqq\supp\bA''$.

For our proof, generation processes in which the support strictly increases, as in the case of $\bA$, are most important.
We express the process by the appending operation as $\bB=\calA^{\hY\hY}_{(2,1)\app(2,0)}(\bA)$ or $\bB=\calA^{\hY\hY\app\hX}_{(2,1)\app(2,0)}(\bA)$.
See \rlb{D1C1E2} for the latter notation.
}
\label{f:AB}
\end{figure}

Let $\bh$ be a product that appears (with a nonzero coefficient) in the Hamiltonian, e.g., $\bh=\hX_u\hX_v$, $\hY_u\hY_v$, or $\hZ_u$ (when $\hz\ne0$).
For $\bA\in\PL$, the commutator $[\bh,\bA]$ is either vanishing or
\eq
[\bh,\bA]=(\text{nonzero constant})\,\bB,
\lb{hAB}
\en
with some $\bB\in\PL$.
When we have \rlb{hAB}, we say $\bB$ is generated by $\bA$ (with $\bh$).
See Figure~\ref{f:AB}.

When $\bB$ is generated by $\bA$, the supports of the two products are related as (i)~$\supp\bB\supsetneqq\supp\bA$, (ii)~$\supp\bB=\supp\bA$, or (iii)~$\supp\bB\subsetneqq\supp\bA$.
It will turn out that case (i), in which the support of $\bB$ is strictly larger than that of $\bA$, plays a central role in our proof.
We shall examine this case in detail and prepare a special symbol $\calA^{\hW\hW}_{u\app v}$.

Clearly, we have $\supp\bB\supsetneqq\supp\bA$ if and only if\footnote{
Here, $\hW_u\hW_v$ means a repeated pair of the same Pauli matrix, e,g, $\hY_u\hY_v$.
In more common expressions like \rlb{A}, on the other hand, $\hA_u$ and $\hA_v$ with $u\ne v$ may independently take any values in $\{\hX_u,\hY_u,\hZ_u\}$ and $\{\hX_v,\hY_v,\hZ_v\}$, respectively.
} $\bh=\hW_u\hW_v$ with $|u-v|=1$, $u\not\in\supp\bA$, $v\in\supp\bA$, and $\hW_v\neq\hA_v$.
Here, $\hW\in\{\hX,\hY,\hZ\}$ if $\Jz\ne0$ and $\hW\in\{\hX,\hY\}$ if $\Jz=0$.
By using \rlb{WW}, the commutator is evaluated as
 \eqa
 [\hW_u\hW_v,\bA]&=\hW_u\otimes[\hW_v,\hA_v]\otimes\biggl(\hsp\bigotimes_{s\in\supp\hA\backslash\{v\}}\hA_s\biggr)
 \nl
 &=2\ri\,\sigma(\hW_v,\hA_v) \,\calA^{\hW\hW}_{u\app v}(\bA).
 \lb{WWA}
 \ena
Here, $\calA^{\hW\hW}_{u\app v}(\bA)\in\PL$ is the product of Pauli matrices defined by 
 \eq
 \calA^{\hW\hW}_{u\app v}(\bA)=
 \hW_u\otimes\dive{\hW_v\hA_v}\otimes\biggl(\hsp\bigotimes_{s\in\supp\hA\backslash\{v\}}\hA_s\biggr).
 \lb{App}
 \en 
We call $ \calA^{\hW\hW}_{u\app v}$ the appending operation. 
It appends to $\bA$ an extra spin operator $\hW$ at  $u$ adjacent to $v\in\supp\bA$ by taking the commutator with $\hW\hW$.
We obvioulsy have $\supp\calA^{\hW\hW}_{u\app v}(\bA)=\supp\bA\cup\{u\}$.

Suppose that $\bB$ is generated by $\bA$ as $\bB=\calA^{\hW\hW}_{u\app v}(\bA)$.
In this case, we also say that $\bA$ is obtained by truncating the site $u$ from $\bB$. 
See Figure~\ref{f:AB}.

It is obvious that $\wid\calA^{\hW\hW}_{u\app v}(\bA)$ is equal to $\wid\bA$ or $\wid\bA+1$.
We therefore have the condition $\wid\bB\,\le\,\km+1$ in the summation in \rlb{HQexp}.

Let us briefly see the remaining two cases.
(ii)~If we have $\bA'\in\PL$ and $\bh=\hW_u$ such that $u\in\supp\bA'$ and $\hW_u\ne\hA'_u$, then the commutator $[\bh,\bA']$ generates $\bB$ with $\supp\bB=\supp\bA'$.
(iii)~If we have $\bA''\in\PL$ and $\bh=\hW_u\hW_v$ with $|u-v|=1$ such that $\{u,v\}\subset\supp\bA''$, $\hW_u=\hA''_u$, and $\hW_v\neq\hA''_v$, then $[\bh,\bA'']$ generates $\bB$ with $\supp\bB=\supp\bA''\backslash\{u\}$.\footnote{
If $\{u,v\}\subset\supp\bA''$,  $\hW_u\neq\hA''_u$, and $\hW_v\neq\hA''_v$, then we have $[\bh,\bA'']=0$ from the anticommutation relations \rlb{XYZ} btween the Pauli matrices.
}
See Figure~\ref{f:AB}.
 
Take an arbitrary $\bB\in\PL$, and  let $\bA^{(1)},\ldots,\bA^{(n)}\in\PL$ be all products that satisfy $\wid\bA^{(j)}\le\km$ and generate $\bB$.
This means that the coefficient $\cB$ in the expansion \rlb{HQexp} is a linear combination of $q_{\bA^{(1)}},\ldots,q_{\bA_n}$, i.e., $\cB=\sum_{j=1}^n\la_j\, q_{\bA^{(j)}}$ (which is a rewriting of \rlb{cB2}).
We then find from \rlb{c=0} that
\eq
\sum_{j=1}^n\la_j\, q_{\bA^{(j)}}=0,
\lb{laq=0}
\en
with nonzero constants $\la_1,\ldots,\la_n$.
We shall use the relation \rlb{laq=0} repeatedly in our proof.
One may determine the constants $\la_j$ explicitly from \rlb{H}, \rlb{Q}, \rlb{HQexp}, and \rlb{WWA}, but such detailed information is not necessary for the moment.  

For $n=1$ and $n=2$, the relation \rlb{laq=0} leads to the following useful lemmas.
\begin{lemma}\label{L:qA=0}
Suppose that there is $\bB\in\PL$ generated by a unique product $\bA\in\PL$ with $\wid\bA\le\km$.
Then we have
\eq
\qA=0.
\en
\end{lemma}
\begin{lemma}\label{L:qA=qA}
Suppose that there is $\bB\in\PL$ generated by exactly two products $\bA,\bA'\in\PL$ with $\wid\bA\le\km$ and $\wid\bA'\le\km$.
Then we have
\eq
\qA=\la\,q_{\bA'},
\lb{qA=qA}
\en
with a nonzero constant $\la$.
\end{lemma}

\subsection{Reduction to a one-dimensional problem}\label{S:reduction}
In the first step of the proof, we show that our problem in $d$-dimensions can be reduced to an essentially one-dimensional problem.
We fix $\km$ such that $2\le\km\le L/2$.
Here, we mainly consider the width in the 1-direction, namely, the horizontal width $\wido\bA$ of a product $\bA$.
We denote by
\eq
\eo=(1,0,\ldots,0),
\lb{eo}
\en
the unit vector in the 1-direction.

Let $\bA\in\PL$ be such that $\wid\bA=\km$.
There is $\alpha\in\{1,\ldots,d\}$ for which $\wid_\alpha\bA=\km$.
Without loss of generality, we assume $\alpha=1$ and hence $\wido\bA=\km$.\footnote{\label{fn:wido}
Our proof uses only the assumption $\wido\bA=\km$ and does not use $\wid_\alpha\bA\le\km$ for $\alpha=2,\ldots,d$.
See Remark~1 at the end of the subsection.
}
Then, by definition there is $a\in\{1,\ldots,L\}$ such that
\eq
0\le(u)_1-a\le\km-1\ ({\rm mod}\ L),
\en
for any $u\in\supp\bA$.
We see $a$ is unique because of the assumption $\km\le L/2$.
We say that $x\in\supp\bA$ is a right-most site of $\bA$ if $(x)_1-a=\km-1$, and, likewise, $y\in\supp\bA$ is a left-most site of $\bA$ if $(y)_1-a=0$.
Note that any $\bA\in\PL$ has at least one right-most site and one left-most site by definition.

\begin{figure}
\centerline{\epsfig{file=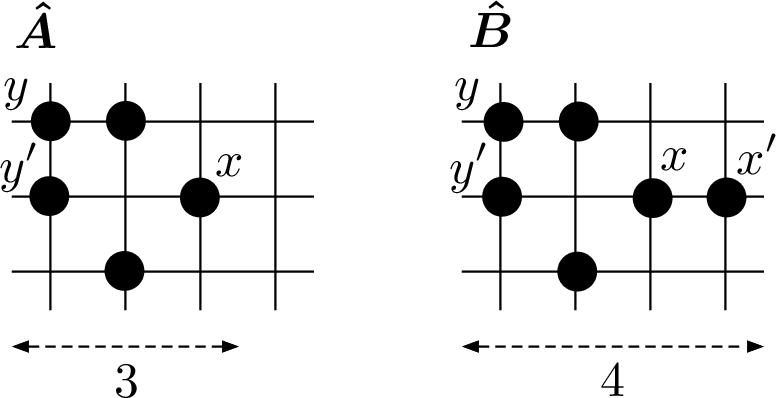,width=6truecm}}
\caption[dummy]{
Black disks represent the Pauli matrices.
Here, the figure represents a small portion of a bigger lattice.
The product $\bA$ with horizontal width $\km=3$ has a right-most site $x$ and two left-most sites $y$ and $y'$.
The product $\bB$ with horizontal width 4 is obtained by appending $x'=x+\eo$ to $\bA$.
One readily sees that $\bA$ is the only product with horizontal width 3 that generates $\bB$.
}
\label{f:nonunique}
\end{figure}

The following lemma highlights a property of the coeffcients $\qA$ that was not encountered in one-dimensional models.
\begin{lemma}\label{L:non-unique}
Let $\bA\in\PL$ be such that $\wido\bA=\km$.
If either right-most sites or left-most sites of $\bA$ are not unique, then we have $\qA=0$.
\end{lemma}
{\em Proof:}\/
Assume, without loss of generality, that left-most sites are not unique.
Let $x$ be a right-most site (which may be unique or not), and $y$ and $y'$ be distinct left-most sites.
We define $\bB= \calA^{\hW\hW}_{x'\app x}(\bA)$ with $x'=x+\eo$ and an appropriate $\hW$.
Clearly, $\wido\bB=\km+1$.
By definition $\bA$ generates $\bB$.
See Figure~\ref{f:nonunique}.

We argue that $\bA$ is the only product with $\wido\le\km$ that generates $\bB$.
To see this, it suffices to note that the truncation of sites $y$ or $y'$ does not reduce the horizontal width.
We then find from Lemma~\ref{L:qA=0} that $\qA=0$.~\qedm

\begin{figure}
\centerline{\epsfig{file=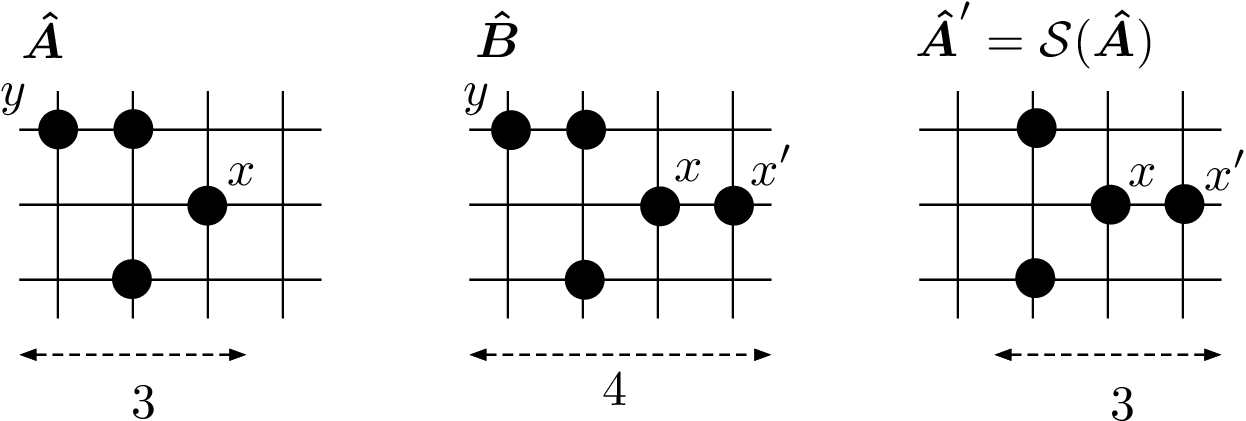,width=9.5truecm}}
\caption[dummy]{
The product $\bA$ with horizontal width $\km=3$ has a unique right-most site $x$ and a unique left-most site $y$.
The product $\bB$ with horizontal width 4 is obtained by appending $x'=x+(1,0)$ to $\bA$.
By truncating the left-most site $y$ from $\bB$ (when possible), we get $\bA'=\calS(\bA)$ with horizontal width 3, which is the shift of $\bA$.
Here, the figure represents a small portion of a bigger lattice.
}
\label{f:shift}
\end{figure}

 \medskip
Following the idea of \cite{Shiraishi2019}, we now introduce the procedure that we call shift, with which we can show the coefficient $\qA$ of any product $\bA$ with $\wido\bA=\km$ is either vanishing or proportional to the coefficients of products in a standard form.
 Our goal is Lemma~\ref{L:1D}.
 Suppose that $\bA\in\PL$ satisfies $\wido\bA=\km$ and has a unique right-most site $x$ and a unique left-most site $y$.
 As in the above proof, we define
 \eq
 \bB= \calA^{\hW\hW}_{x'\app x}(\bA),
 \en
 with $x'=x+\eo$, which has $\wido\bB=\km+1$.
 See Figure~\ref{f:shift}.
 Here, we chose $\hW$ according to the convention
 \eq
 \hW=
 \begin{cases}
 \hX,&\text{if $\hA_x=\hY_x$ or $\hZ_x$};\\
 \hY,&\text{if $\hA_x=\hX_x$},
 \end{cases}
 \en
no matter whether we are treating the XY or XYZ model.
We then ask if there is $\bA'\in\PL$, other than $\bA$, such that $\wido\bA'\le\km$ and generates $\bB$.
Since the horizontal width of $\bA'$ is strictly smaller than that of $\bB$, we see that $\bB$ must be generated by $\bA'$ through an appending operation that appends the left-most site $y$ as 
 \eq
 \bB=\calA^{\hW'\hW'}_{y\app y+\eo}(\bA'),
 \lb{B=AWWA'}
 \en
 with some $\hW'\in\{\hX,\hY,\hZ\}$ if $\Jz\ne0$ or $\hW'\in\{\hX,\hY\}$ if $\Jz=0$.
 Here, it is necessary (but not sufficient) that $y+\eo\in\supp\bA'$.
 In other words, $\bA'$ is obtained from $\bB$ by truncating the left-most site $y\in\supp\bB$.
 If such $\bA'$ exists, we call it the shift of $\bA$ and write $\calS(\bA)=\bA'$.
 It is crucial that the shift $\calS(\bA)$ is unique if it exists.
This is because the unique left-most site $y$ is neighboring at most one site in $\supp\bB$ and hence can be truncated in (at most) one manner.\footnote{
In general, a given site in the support of a product can be truncated in multple manners.
In the examples given in Figure~\ref{f:AB}, the product $\bB$ is also generated by $\hZ_{(1,1)}\hZ_{(2,0)}\hX_{(4,0)}$ with $\hY_{(1,1)}\hY_{(2,1)}$.
We see $\hY_{(2,1)}$ in $\bB$ can be truncated in two different ways.
 }
 See Figure~\ref{f:shift}.
 When the shift $\calS(\bA)$ exists, we see that $\bA$ and $\bA'=\calS(\bA)$ are the only products with $\wido\le\km$ that generate $\bB$.
Then we see from Lemma~\ref{L:qA=qA} that $\qA$ and $q_{\bA'}$ are related by \rlb{qA=qA}.
 
If there exist no $\bA'$ with $\wido\bA'\le\km$ satisfying \rlb{B=AWWA'}, we say that the shift $\calS(\bA)$ does not exist.
In this case, we see from Lemma~\ref{L:qA=0} that $\qA=0$.
We also say that the shift $\calS(\bA)$ does not exist when $\bA\in\PL$ with $\wido\bA=\km$ has non-unique left-most sites or non-unique right-most sites.
 We have seen in Lemma~\ref{L:non-unique} that $\qA=0$ in such a case.

To summarize, we have shown the following.
\begin{lemma}\label{L:inheritance}
Let $\bA\in\PL$ be any product of Pauli matrices such that $\wido\bA=\km$.
If the shift $\calS(\bA)$ exists then
\eq
\qA=\la\,q_{\calS(\bA)},
\en
with a nonzero constant $\la$.
If the shift $\calS(\bA)$ does not exist, then $\qA=0$.
\end{lemma}
 
 Let us state a necessary condition that a product $\bA\in\PL$ can be shifted $\km$ times or more.
This condition will play an essential role in the rest of the proof.
\begin{lemma}\label{L:manyshifts}
Let $\bA\in\PL$ be any product of Pauli matrices such that $\wido\bA=\km$.
Its $n$-fold shift $\calS^n(\bA)=\underbrace{\calS\circ\cdots\circ\calS}_{n}(\bA)$ with $n\ge\km$ exists only when $\bA$ is a product of Pauli operators on a continuous horizontal segment with $\km$ sites, i.e., only when $\wid_\alpha(\bA)=1$ for $\alpha=2,\ldots,d$ and $\supp\bA$ consists of $\km$ sites.
If the condition is not satisfied, then we have $\qA=0$ from Lemma~\ref{L:inheritance}.
We see, in particular, that any $\bA\in\PL$ such that $\wido\bA=\km$ and $\wid_\alpha\bA\ge2$ for some $\alpha=2,\ldots,d$ has $\qA=0$.
\end{lemma}
\noindent{\em Proof:}\/
Recall that in order for the left-most site $y$ in $\bA$ (which is also the left-most site of $\bB= \calA^{\hW\hW}_{x'\app x}(\bA)$) to be truncatable, it must be that $y+\eo\in\supp\bA$.
Using this observation repeatedly, we see that $\bA$ must contain Pauli matrices on a continuous horizontal segment with $\km$ sites.
Any Pauli matrix at a site out of the horizontal line is excluded since it leads to non-unique left-most sites in $\calS^m(\bA)$ for some $m<\km$.~\qedm
\medskip

We see that the multiple shift $\calS^n(\bA)$, if exists, settles to standard forms after sufficiently many shifts.
\begin{lemma}\label{L:SnA}
Let $\bA\in\PL$ be any product of Pauli matrices such that $\wido\bA=\km$ and $\calS^{\km}(\bA)$ exists.
Then, for $n=\km-1$ or $\km$, the $n$-fold shift $\calS^n(\bA)$ is identical to
\eq
\bC_{\rm XX}=\hX_{x_0+\eo}\otimes\biggl(\hsp\bigotimes_{j=2}^{\km-1}\hZ_{x_0+j\eo}\biggr)\otimes\hX_{x_0+\km\eo},
\lb{CXX}
\en
or
\eq
\bC_{\rm YX}=\hY_{x_0+\eo}\otimes\biggl(\hsp\bigotimes_{j=2}^{\km-1}\hZ_{x_0+j\eo}\biggr)\otimes\hX_{x_0+\km\eo},
\lb{CYX}
\en
for a suitable $x_0\in\La$.
\end{lemma}
\noindent{\em Proof:}\/
Suppose that $\bA''=\calS^{\km-1}(\bA)$ exists.
Since we add $\hX$ or $\hY$ to the right in the appending process, the right-most site of $\bA''$ has $\hX$ or $\hY$.
Examining the process of repeated shifts, one also finds from \rlb{App} and $|\hX\hY|=|\hY\hX|=\hZ$ that all sites in $\supp\bA''$ except for the right-most or left-most sites have $\hZ$.
We can also assume $\supp\bA''$ consists of $\km$ sites aligned contiguously in the 1-direction since, otherwise, $\calS(\bA'')$ does not exist.

We note that $\calS(\bA'')$ does not exist if the left-most site of $\bA''$ has $\hZ$.
This is because one cannot truncate the two contiguous $\hZ$'s at the left end of $\bA''$.
(In other words, no appending operations produce two contiguous $\hZ$'s at the end.)
Thus the left-most site of $\bA''$ has either $\hX$ or $\hY$.
If the right-most site of $\bA''$ has $\hX$, we see $\hA''$ equals either \rlb{CXX} or \rlb{CYX}.
If this is not the case, we see that the right-most site of $\calS(\bA'')$ has $\hX$, and $\calS(\bA'')$ equals either \rlb{CXX} or \rlb{CYX}.~\qedm

\medskip
From Lemmas~\ref{L:inheritance} and \ref{L:SnA}, we readily arrive at the following conclusion of the present subsection.
As promised, we have shown that it suffices to examine products of Pauli matrices that lie on a one-dimensional subset.
\begin{lemma}\label{L:1D}
Let $\bA\in\PL$ be any product of Pauli matrices such that $\wido\bA=\km$.
Then we have either
\eq
\qA=\la'\,q_{\bC_{\rm XX}},\quad
\qA=\la''\,q_{\bC_{\rm YX}},\quad\text{or}\quad
\qA=0,
\en
for a suitable $x_0$ and nonzero $\la'$ and $\la''$.
\end{lemma}

In Sections~\ref{S:k=3}, \ref{S:preliminaries}, \ref{S:CYX}, and \ref{S:CXX}, we shall treat the case with $3\le\km\le L/2$ and prove that $q_{\bC_{\rm XX}}=q_{\bC_{\rm YX}}=0$.
By Lemma~\ref{L:1D}, this implies $\qA=0$ for any $\bA\in\PL$ such that $\wido\bA=\km$.
Since there is nothing special about the 1-direction, this implies $\qA=0$ for any $\bA\in\PL$ such that $\wid\bA=\km$.
As we discussed in the first paragraph of Section~\ref{s:basic}, this proves our main result, Theorem~\ref{T:main}.

\medskip\noindent
{\bf Remarks:}
\par\noindent
1. As we noted in footnote~\ref{fn:wido}, all the results in the present subsection follow only from the assumption $\wido\bA\le\km$.
Since there is nothing special about the 1-direction, it is enough to assume that the width of $\bA$ in one arbitrary direction, say $\alpha$, rather than all the directions, does not exceed $\km$.
We can therefore prove a stronger statement than Theorem~\ref{T:main}, namely, for each $\alpha=1,\ldots,d$ and $3\le\km\le L/2$, there exists no conserved quantity of the form $\hQ=\sum_{\bA\in\PL\,(\wid_\alpha\bA\le\km)}\qA\bA$ with $\qA\ne0$ for some $\bA$ with $\wid_\alpha\bA=\km$.

One may interpret our proof of the absence of nontrivial local conserved quantities as that for a one-dimensional model in which each site carries a ``spin'' that consists of $L^{d-1}$ spins on the corresponding hyperplane.

\medskip\par\noindent
2. The condition stated in Lemma~\ref{L:manyshifts} is far from sufficient.
If $\Jz\ne0$, we can prove a stronger condition parallel to Lemma~1 in Section~3.1 of \cite{Shiraishi2024}.
But such detailed information is not necessary for our purpose.
\medskip\par\noindent
3. Note that one can write\footnote{
The divestment operation $|\cdots|$ is defined for a single spin in \rlb{divest}.
In a system of multiple sites, it operates on each site separately as
$|(\hX_{x_1}\hX_{x_2})(\hY_{x_2}\hY_{x_3})|=|\hX_{x_1}|\,|\hX_{x_2}\hY_{x_2}|\,|\hY_{x_3}|=\hX_{x_1}\hZ_{x_2}\hY_{x_3}$.
}
\eq
\bC_{\rm XX}=\dive{
(\hX_{x_1}\hX_{x_2})(\hY_{x_2}\hY_{x_3})(\hX_{x_3}\hX_{x_4})\ldots(\hX_{x_{\km-1}}\hX_{x_{\km}})
},
\lb{DB1}
\en
when $\km$ is even, and 
\eq
\bC_{\rm YX}=\dive{
(\hY_{x_1}\hY_{x_2})(\hX_{x_2}\hX_{x_3})(\hY_{x_3}\hY_{x_4})\ldots(\hX_{x_{\km-1}}\hX_{x_{\km}})
},
\lb{DB2}
\en
when $\km$ is odd, where $x_j=x_0+j\eo$.
For the reader familiar with \cite{Shiraishi2024}, we note that these are nothing but the doubling products $\overline{\hX\hY\hX\hY\cdots\hY\hX}_{x_{\km}}$ and $\overline{\hY\hX\hY\hX\cdots\hY\hX}_{x_{\km}}$, respectively.
On the other hand, $\bC_{\rm XX}$ for odd $\km$ or $\bC_{\rm YX}$ for even $\km$ does not have such a doubling product expression.\footnote{
If $\Jz\ne0$, one can use the argument in \cite{Shiraishi2024} to show that these products have vanishing coefficients.
We do not use this argument in the present proof.
}

\subsection{The case with $\km=3$}\label{S:k=3}
From now on, we shall assume $3\le\km\le L/2$ and make use of the high-dimensional nature of the problem to prove $q_{\bC_{\rm XX}}=q_{\bC_{\rm YX}}=0$.
We work in two dimensions for notational simplicity, but we do not lose generality.
One can always interpret the coordinate, say, $(1,2)$ as the abbreviation of $(1,2,0,\ldots,0)$ to recover the case with general $d\ge2$.

\begin{figure}
\centerline{\epsfig{file=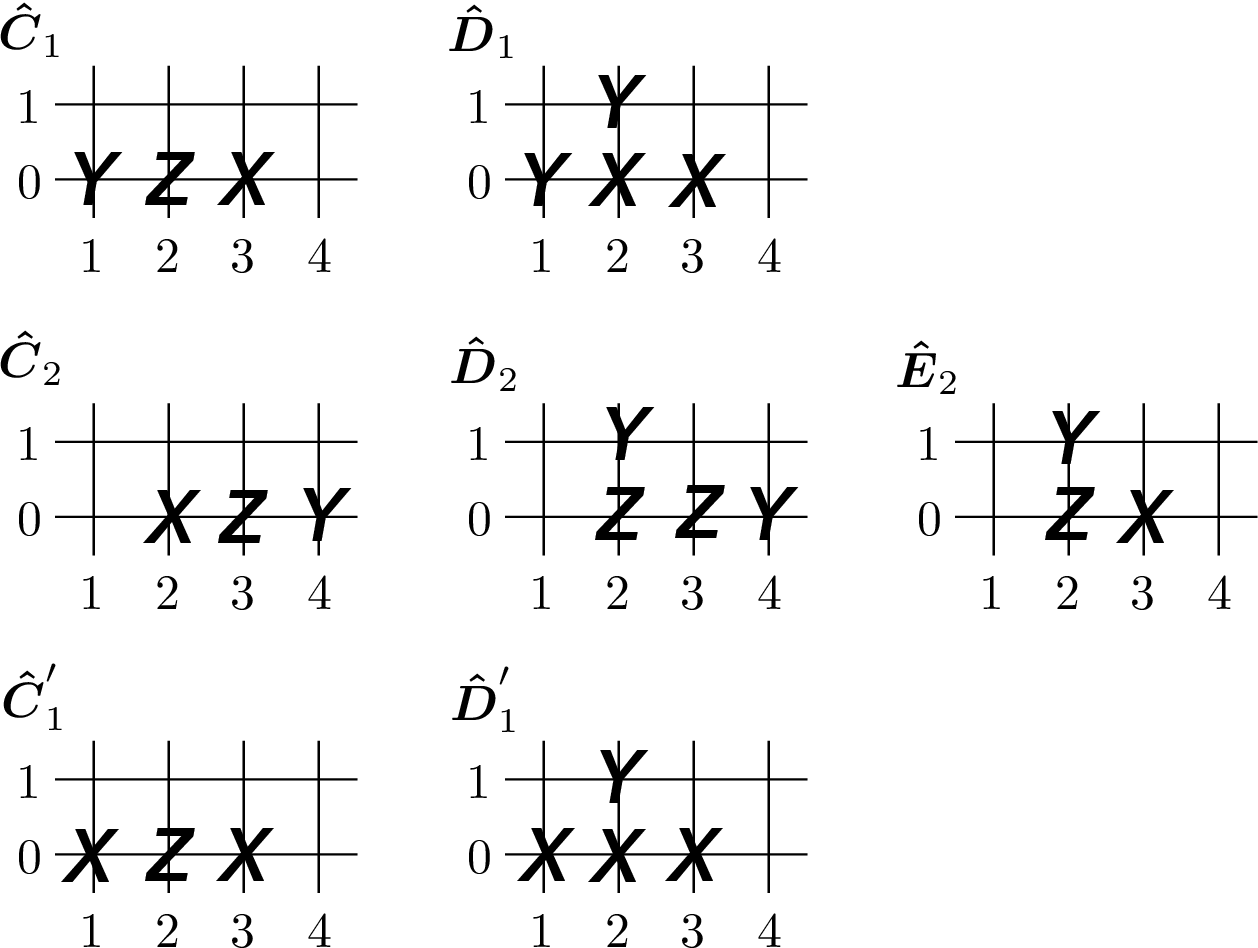,width=8truecm}}
\caption[dummy]{
The products of Pauli operators that play central roles in Section~\ref{S:k=3}.
Here, the figure represents a small portion of a bigger lattice.
}
\label{f:CDE3}
\end{figure}

In this subsection, we study the simplest case with $\km=3$, where we already see the essence of the general proof.
We may assume $x_0=(0,0)$ by translating the coordinate.  Note that we are not assuming the translational invariance.
Then, $\bC_{\rm YX}$, defined in \rlb{CYX}, becomes
\eq
\bC_1=\hY_{(1,0)}\hZ_{(2,0)}\hX_{(3,0)},
\en
where we introduced the new symbol $\bC_1$ for simplicity.
Our goal is to show $q_{\bC_1}=0$.
We also define
\eq
\bD_1=\hY_{(1,0)}\hY_{(2,1)}\hX_{(2,0)}\hX_{(3,0)},\quad
\bE_2=\hY_{(2,1)}\hZ_{(2,0)}\hX_{(3,0)}.
\en
See Figure~\ref{f:CDE3}.
Note that $\wido\bD_1=3=\km$ and $\wid_2\bD_1=2$.

It is readily confirmed that $\bC_1$ and $\bE_2$ generate $\bD_1$.
More precisely, we have
\eq
\bD_1=\calA^{\hY\hY\app\hZ}_{(2,1)\app(2,0)}(\bC_1)
=\calA^{\hY\hY\app\hZ}_{(1,0)\app(2,0)}(\bE_2).
\lb{D1C1E2}
\en
Here, we have rewritten the symbol $\calA^{\hW\hW}_{u\app v}$ for the appending operation as $\calA^{\hW\hW\app\hA_v}_{u\app v}$ to indicate the Pauli matrices involved in the process explicitly.
See \rlb{App}.
This seemingly cumbersome notation is convenient for bookkeeping the coefficients $\la_j$ in the basic relation \rlb{laq=0}.

The relations \rlb{D1C1E2} say that $\bC_1$ and $\bE_2$, respectively, are obtained by trncating the branch $(2,1)$ and the left-most site $(1,0)$, respectively, from $\bD_1$.
It is crucial that one cannot truncate the right-most site $(3,0)$ from $\bD_1$.
This is because any operation that appends $(3,0)$ cannot generate a repeated pair like $\hX_{(2,0)}\hX_{(3,0)}$ that appears in $\bD_1$.
The same observation will be used in Sections~\ref{S:CYX} and \ref{S:CXX}.  

Of course, $\bC_1$ and $\bE_2$ are not the only products that generate $\bD_1$.
However, thanks to Lemma~\ref{L:manyshifts}, we don't need to take into account other products, as we shall see now.
There is one more product $\bC_1''=\hZ_{(1,0)}\hY_{(2,1)}\hX_{(3,0)}$ that genereates $\bD_1$ by an appending operation as $\bD_1=\calA^{\hX\hX\app\hZ}_{(2,0)\app(1,0)}(\bC_1'')$.
There are also products that generate $\bD_1$ with magnetic filed parts of the Hamiltonian.
These products have the same support as $\bD_1$.
Note that all these products, including $\bC_1''$, have $\wido=3=\km$ and $\wid_2=2$.
We then see from Lemma~\ref{L:manyshifts} that their coefficients $q$ are zero.
There are also products that generate $\bD_1$ and have strictly larger support than $\bD_1$.
They also have zero coefficients for the same reason.

We conclude that we only need to consider $\bC_1$ and $\bE_2$ in the basic relation \rlb{laq=0}.
Now, by using the relation \rlb{WWA} between the commutator and the appending operation, we see that the coefficient of $c_{\bD_1}$ defined by  \rlb{HQexp} is given by
\eq
c_{\bD_1}=2\ri(\Jy\,q_{\bC_1}+\Jy\,q_{\bE_2}).
\en
Since $c_{\bD_1}=0$ (as in \rlb{c=0}), we find
\eq
q_{\bC_1}+q_{\bE_2}=0.
\lb{3qq1}
\en

To get further information, we define
\eq
\bC_2=\calS(\bC_1)=\hX_{(2,0)}\hZ_{(3,0)}\hY_{(4,0)},\quad
\bD_2=\calS(\bD_1)=\hY_{(2,1)}\hZ_{(2,0)}\hZ_{(3,0)}\hY_{(4,0)}.
\en
We have 
\eq
\bD_2=\calA^{\hY\hY\app\hX}_{(2,1)\app(2,0)}(\bC_2)=\calA^{\hY\hY\app\hX}_{(4,0)\app(3,0)}(\bE_2).
\en
We again see that $\bC_2$ and $\bE_2$ are the only products that generate $\bD_2$ and are not excluded by Lemma~\ref{L:manyshifts}.
Repeating the same procedure as above, we find
\eq
q_{\bC_2}+q_{\bE_2}=0.
\lb{3qq2}
\en

We now claim that the coefficients of $\bC_1$ and its shift $\bC_2=\calS(\bC_1)$ are related by
\eq
q_{\bC_1}+q_{\bC_2}=0.
\lb{3qq3}
\en
This is easily shown by considering the product $\hY_{(1,0)}\hZ_{(2,0)}\hZ_{(3,0)}\hY_{(4,0)}$, but we shall leave the proof to Section~\ref{S:preliminaries}.
See \rlb{qCj}.
Combining \rlb{3qq1}, \rlb{3qq2}, and \rlb{3qq3}, we arrive at the desired result $q_{\bC_1}=0$.

In the same setting with $\km=3$, the product $\bC_{\rm XX}$, defined in \rlb{CXX}, becomes
\eq
\bC'_1=\hX_{(1,0)}\hZ_{(2,0)}\hX_{(3,0)}.
\en
See, again, Figure~\ref{f:CDE3}.
This is easier to treat.
Let
\eq
\bD_1'=\calA^{\hY\hY\app\hZ}_{(2,1)\app(2,0}(\bC_1')
=\hX_{(1,0)}\hY_{(2,1)}\hX_{(2,0)}\hX_{(3,0)}.
\en
Here, we see that $\bC_1'$ is the only product with $\wido\le3=\km$ that generates $\bD_1'$ and not excluded by Lemma~\ref{L:manyshifts}.
Thus we get $q_{\bC_1'}=0$ from Lemma~\ref{L:qA=0}.
This concludes the proof of Theorem~\ref{T:main} in the case $\km=3$.

\subsection{Preliminaries for general $\km\ge3$}\label{S:preliminaries}
In Sections~\ref{S:preliminaries}, \ref{S:CYX}, and \ref{S:CXX}, we focus on the general case with $3\le\km\le L/2$ and prove that $q_{\bC_{\rm YX}}=q_{\bC_{\rm XX}}=0$.
As was discussed at the end of Section~\ref{S:reduction}, this proves our main result, Theorem~\ref{T:main}.
In what follows, we abbreviate $\km$ as $\kn$ to shorten equations.

As in Section~\ref{S:k=3}, we note that the products $\bC_{\rm YX}$ and $\bC_{\rm XX}$, defined in \rlb{CYX} and \rlb{CXX}, become
\eq
\bC_1=\hY_{(1,0)}\otimes\ZZ{2}{\kn-1}\otimes\hX_{(\kn,0)},
\lb{kC1}
\en
and
\eq
\bC'_1=\hX_{(1,0)}\otimes\ZZ{2}{\kn-1}\otimes\hX_{(\kn,0)},
\lb{kC'1}
\en
respectively, after suitable translation.
We again introduced the simpler symbols $\bC_1$ and $\bC'_1$.
Let us also write
\eq
q=q_{\bC_1},\quad q'=q_{\bC'_1}.
\en
Our goal is to show $q=q'=0$.

For $j=1,2,\ldots$, we deifine the shifted products $\bC_j=\calS^{j-1}(\bC_1)$ and $\bC'_j=\calS^{j-1}(\bC'_1)$.
More specifically,
\eq
\bC_j=
\begin{cases}
\displaystyle
\hY_{(j,0)}\otimes\ZZ{j+1}{j+\kn-2}\otimes\hX_{(j+\kn-1,0)},&\jodd;\\
\displaystyle
\hX_{(j,0)}\otimes\ZZ{j+1}{j+\kn-2}\otimes\hY_{(j+\kn-1,0)},&\jeven,
\end{cases}
\lb{Cj}
\en
\eq
\bC'_j=
\begin{cases}
\displaystyle
\hX_{(j,0)}\otimes\ZZ{j+1}{j+\kn-2}\otimes\hX_{(j+\kn-1,0)},&\jodd;\\
\displaystyle
\hY_{(j,0)}\otimes\ZZ{j+1}{j+\kn-2}\otimes\hY_{(j+\kn-1,0)},&\jeven.
\end{cases}
\lb{C'j}
\en
We shall show that the coefficients for these products satisfy
\eq
q_{\bC_j}=\begin{cases}
q,&\jodd;\\
-q,&\jeven,
\end{cases}
\lb{qCj}
\en
and
\eq
q_{\bC'_j}=\begin{cases}
q',&\jodd;\\
\kappa\,q',&\jeven,
\end{cases}
\lb{qC'j}
\en
where we introduced the anisotropy parameter
\eq
\kappa=\frac{\Jy}{\Jx}\ne0.
\lb{kappa}
\en
Note that \rlb{3qq3} in Section~\ref{S:k=3} is a special case of \rlb{qCj}.

To prove \rlb{qC'j}, we assume $j$ is even and define
\eq
\bF'_{j-1}=\hX_{(j-1,0)}\otimes\ZZ{j}{j+\kn-2}\otimes\hY_{(j+\kn-1,0)},
\en
which has $\wido\bF'_{j-1}=\kn+1$.
We then see that
\eq
\bF'_{j-1}=\AXY_{(j-1,0)\app(j,0)}(\bC'_j)=\AYX_{(j+\kn-1,0)\app(j+\kn-2,0)}(\bC'_{j-1}),
\en
and that $\bC'_j$ and $\bC'_{j-1}$ are the only products with $\wido\le\kn$ that generate $\bF'_{j-1}$.
This implies
\eq
c_{\bF'_{j-1}}=2\ri\{\Jx\,q_{\bC'_j}-\Jy\,q_{\bC'_{j-1}}\},
\lb{cF'}
\en
where $c_{\bF'_{j-1}}$ is defined by \rlb{HQexp}.
Here, we used \rlb{WWA} and \rlb{sign} to determine the exact coefficients in the right-hand side.
Since we have $c_{\bF'_{j-1}}=0$ as in \rlb{c=0}, we arrive at
\eq
q_{\bC'_j}=\kappa\,q_{\bC'_{j-1}}\quad\text{(even $j$)}.
\lb{qC'jeven}
\en
Repeating the same argument with
\eq
\bF'_{j}=\hY_{(j,0)}\otimes\ZZ{j+1}{j+\kn-1}\otimes\hX_{(j+\kn,0)},
\en
we get
\eq
q_{\bC'_j}=\kappa\,q_{\bC'_{j+1}}\quad\text{(even $j$}),
\en
which, with \rlb{qC'jeven}, implies the desired \rlb{qC'j}.
The relation \rlb{qCj} is proved in a similar (indeed, easier) manner.

\begin{figure}
\centerline{\epsfig{file=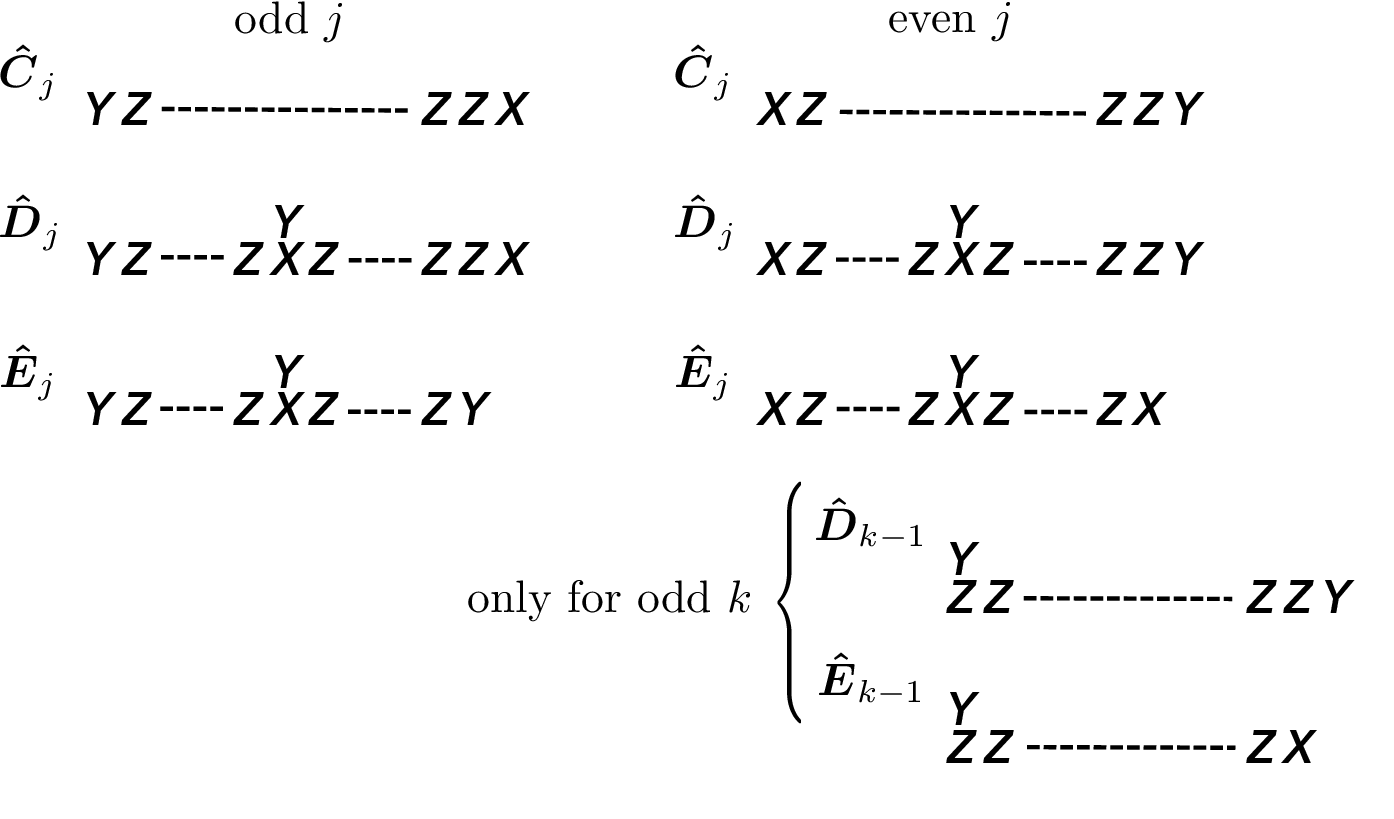,width=10truecm}}
\caption[dummy]{
The products $\bC_j$, $\bD_j$, and $\bE_j$ that play essential role in Section~\ref{S:CYX}.
The dashed lines represent repeated $\hZ$.
Note that $\bE_1$ is not defined.
When $\km$ (which is denoted simply as $\kn$ for simplicity in Sections~\ref{S:preliminaries}, \ref{S:CYX}, and \ref{S:CXX}) is odd, then the products $\bD_j$, and $\bE_j$ with the largest $j$, which is $\kn-1$, are exceptional and should be treated separately.
}
\label{f:CDEk}
\end{figure}

\subsection{The treatment of $\bC_{\rm YX}$}\label{S:CYX}
Let us examine the products $\bC_j$ and show that $q=q_{\bC_1}=0$.
Here, the range of $j$ is
\eq
j=\begin{cases}
1,\ldots,\kn-1,&\text{if $\kn$ is odd};\\
1,\ldots,\kn-2,&\text{if $\kn$ is even}.
\end{cases}
\lb{jrange}
\en
As in Section~\ref{S:k=3}, we introduce two more classes of products:
\eqg
\bD_j=\AYZ\kk(\bC_j),\quad\text{for $j\ne k-1$},
\lb{Dj}\\
\bD_{\kn-1}=\AYX\kk(\bC_{\kn-1}),\quad\text{when $\kn$ is odd},
\lb{Dk-1}\\
\bE_j=
\begin{cases}
\displaystyle
\AYZ\kk\biggl(\hY_{(j,0)}\otimes\ZZ{j+1}{j+\kn-3}\otimes\hY_{(j+\kn-2,0)}\biggr),&\text{$j$ odd and $j\ne1$};\\
\displaystyle
\AYZ\kk\biggl(\hX_{(j,0)}\otimes\ZZ{j+1}{j+\kn-3}\otimes\hX_{(j+\kn-2,0)}\biggr),&\text{$j$ even and $j\ne\kn-1$},
\end{cases}
\lb{Ej}\\
\bE_{\kn-1}=\AYX\kk\biggl(\hX_{(\kn-1,0)}\otimes\ZZ{\kn}{2\kn-4}\otimes\hX_{(2\kn-3,0)}\biggr),
\quad\text{when $\kn$ is odd}.
\lb{Ek-1}
\eng
We do not define $\bE_1$.
Note that $\wido\bC_j=\wido\bD_j=\kn$ and $\wido\bE_j=\kn-1$.
It is crucial that $\bD_j$ and $\bE_j$ have a branch at $(\kn-1,1)$, and $\wid_2\bD_j=\wid_2\bE_j=2$.
It is also worth noting that the arguments of the appending operations for $\bE_j$ in \rlb{Ej} and \rlb{Ek-1} are obtained by truncating the right-most site in $\bC_j$.
The reason that the cases with $j=\kn-1$ for odd $\kn$ require separate definitions \rlb{Dk-1}, \rlb{Ek-1} may be read off from Figure~\ref{f:CDEk}.

As in Section~\ref{S:k=3}, we shall consider products that generate $\bD_j$ and deduce conditions for the coefficients of $\bC_j$ and $\bE_j$.

Let us first examine $\bD_j$ with odd $j$.
Observe, for odd $j$ such that $j\ne\kn-2$ when $\kn$ is odd, and for all odd $j$ when $\kn$ is even, that
\eq
\bD_j=\AYZ\kk(\bC_j)=\AXY_{(j+\kn-1,0)\app(j+\kn-2,0)}(\bE_j)=\AYX_{(j,0)\app(j+1,0)}(\bE_{j+1}).
\lb{Dgen1}
\en
In other words, $\bC_j$, $\bE_j$, and $\bE_{j+1}$, respectively, are obtained by truncating the branch $(\kn-1,1)$, the right-most site $(j+\kn-1,0)$, and the left-most site $(j,0)$, respectively, from $\bD_j$.
Exceptionally, one cannot truncate the right-most site from $\bD_1$ for the same reason as discussed in the paragraph after that contains \rlb{D1C1E2}.
We, therefore, ignore the term including $\bE_j$ in \rlb{Dgen1} when $j=1$.

We see $\bD_j$ is generated by $\bC_j$, $\bE_j$, and $\bE_{j+1}$, or, when $j=1$, by $\bC_1$ and $\bE_2$.
There are many other products of Pauli matrices that generate $\bD_j$, but all of them have $\wido\ge\kn$ and $\wid_2\ge2$.
To see this, it is enough to note that $\wido\bD_j=\kn$, $\wid_2\bD_j=2$, and the only ways to reduce one of the widths are to truncate the branch, the right-most site, or the left-most site.
Exactly as in Section~\ref{S:k=3}, we see that all those products that generate $\bD_j$ other than $\bC_j$, $\bE_j$, or $\bE_{j+1}$ have zero coefficients because of Lemma~\ref{L:manyshifts}.
Then, the remainder is, in principle, straightforward.
As we did in \rlb{cF'}, we evaluate the commutators and use the basic condition \rlb{c=0} to conclude
\eq
\kappa\,q_{\bC_j}+q_{\bE_j}-\kappa\,q_{\bE_{j+1}}=0,
\en
which, with \rlb{qCj}, leads to
\eq
\kappa\,q+\tq_j-\kappa\,\tq_{j+1}=0\quad(\text{odd $j$ s.t. $j\ne\kn-2$}),
\lb{R1}
\en
where we wrote for simplicity
\eq
\tq_j=q_{\bE_j}.
\en
Although the case $j=1$ is exceptional, we can use \rlb{R1} as it is by setting
\eq
\tq_1=0.
\lb{tq1=0}
\en
Finally, when $\kn$ is odd, we have
\eq
\bD_{\kn-2}=\AYZ\kk(\bC_{\kn-2})=\AXY_{(2\kn-3,0)\app(2\kn-4,0)}(\bE_{\kn-2})=\AYZ_{(\kn-2,0)\app(\kn-1,0)}(\bE_{\kn-1}).
\lb{Dgen2}
\en
Compared with \rlb{Dgen1}, $\AYX$ in the right-most side is replaced by $\AYZ$, reflecting the definition \rlb{Ek-1} of $\bE_{\kn-1}$.
See Figure~\ref{f:CDEk}.
As above, this leads
\eq
\kappa\,q_{\bC_{\kn-2}}+q_{\bE_{\kn-2}}+\kappa\,q_{\bE_{\kn-1}}=0,
\en
or, equivalently,
\eq
\kappa\,q+\tq_{\kn-2}+\kappa\,\tq_{\kn-1}=0.
\lb{R2}
\en

We next examine $\bD_j$ with even $j$.
We see, for even $j$ such that $j\ne\kn-1$ when $\kn$ is odd, and for all even $j$ when $\kn$ is even,
\eq
\bD_j=\AYZ\kk(\bC_j)=\AYX_{(j+\kn-1,0)\app(j+\kn-2,0)}(\bE_j)=\AXY_{(j,0)\app(j+1,0)}(\bE_{j+1}),
\lb{Dgen3}
\en
which leads to
\eq
\kappa\,q_{\bC_j}-\kappa\,q_{\bE_j}+q_{\bE_{j+1}}=0.
\en
With \rlb{qCj}, this implies
\eq
\kappa\,q+\kappa\,\tq_j-\tq_{j+1}=0\quad(\text{even $j$ s.t. $j\ne\kn-1$}).
\lb{R3}
\en
When $\kn$ is even and $j=\kn-2$, we understand $\tq_{\kn-1}=0$ in \rlb{R3} since $\bD_{\kn-2}$ is generated only by $\bC_{\kn-2}$ and $\bE_{\kn-2}$, as well as other products that are excluded by Lemma~\ref{L:manyshifts}.\footnote{
\label{fn:Dmax}
This is because the left-most operator of $\bD_{\kn-2}$, which is $\hX_{(\kn-2,0)}$, is adjacent to $\hX_{(\kn-1,0)}$ and cannot be truncated.
}
Finally, when $\kn$ is odd, we have
\eq
\bD_{\kn-1}=\AYX\kk(\bC_{\kn-1})=\AYX_{(2\kn-2,0)\app(2\kn-3,0)}(\bE_{\kn-1}),
\en
which leads to the terminal relation
\eq
q_{\bC_{\kn-1}}+q_{\bE_{\kn-1}}=0,
\en
or, equivalently,
\eq
q=\tq_{\kn-1}.
\lb{R4}
\en

The remaining task is to examine the conditions \rlb{R1}, \rlb{R2}, \rlb{R3}, and \rlb{R4} to deduce the desired result $q=0$.
When $\kn$ is even, we only need to sum up \rlb{R1} and \rlb{R3} from $j=1$ to $\kn-2$ to get
\eq
(\kn-2)\kappa\,q+\tq_1-\tq_{\kn-1}=0,
\lb{cond1}
\en
which reduces to $(\kn-2)\kappa\,q=0$ because $\tq_1=\tq_{\kn-1}=0$.
Since $\kn\ge3$, we conclude $q=0$.
When $\kn$ is odd, we sum up \rlb{R1} and \rlb{R3} from $j=1$ to $\kn-3$ to get
\eq
(\kn-3)\kappa\,q+\tq_1-\kappa\,\tq_{\kn-2}=0.
\en
Then we note that \rlb{R2} and \rlb{R4}, which are the relations for $j=\kn-2$ and $\kn-1$, lead to $\tq_{\kn-2}=-2\kappa\,q$.
Again recalling $\tq_1=0$, we find
\eq
(\kn-1)\kappa\,q=0,
\lb{k-1kappa}
\en
which implies $q=0$.

\subsection{The treatment of $\bC_{\rm XX}$}\label{S:CXX}
We shall examine the products $\bC'_j$ and show that $q'=q_{\bC'_1}=0$.
Since the analysis is parallel to that in Section~\ref{S:CYX}, we shall describe only the essential definitions and relations.
For $j$ in the range
\eq
j=\begin{cases}
1,\ldots,\kn-2,&\text{if $\kn$ is odd};\\
1,\ldots,\kn-1,&\text{if $\kn$ is even},
\end{cases}
\lb{jrange'}
\en
we define the following products of Pauli matrices:
\eqg
\bD'_j=\AYZ\kk(\bC'_j),\quad\text{for $j\ne\kn-1$},
\lb{D'j}\\
\bD'_{\kn-1}=\AYX\kk(\bC'_{\kn-1}),\quad\text{when $\kn$ is even},
\lb{D'k-1}\\
\bE'_j=
\begin{cases}
\displaystyle
\AYZ\kk\biggl(\hX_{(j,0)}\otimes\ZZ{j+1}{j+\kn-3}\otimes\hY_{(j+\kn-2,0)}\biggr),&\text{$j$ odd and $j\ne1,\kn-1$};\\
\displaystyle
\AYZ\kk\biggl(\hY_{(j,0)}\otimes\ZZ{j+1}{j+\kn-3}\otimes\hX_{(j+\kn-2,0)}\biggr),&\text{$j$ even},
\end{cases}
\lb{E'j}\\
\bE'_{\kn-1}=\AYX\kk\biggl(\hX_{(\kn-1,0)}\otimes\ZZ{\kn}{2\kn-4}\otimes\hY_{(2\kn-3,0)}\biggr),
\quad\text{when $\kn$ is even}.
\lb{E'k-1}
\eng

For an odd $j$ such that $j\ne\kn-1$, we observe
\eq
\bD'_j=\AYZ\kk(\bC'_j)=\AXY_{(j+\kn-1,0)\app(j+\kn-2,0)}(\bE'_j)=\AXY_{(j,0)\app(j+1,0)}(\bE'_{j+1}),
\lb{D'gen1}
\en
which, after using \rlb{qC'j} and defining $\tq'_j=q_{\bE'_j}$, implies
\eq
\kappa\,q'+\tq'_j+\tq'_{j+1}=0\quad(\text{odd $j$ s.t. $j\ne\kn-1$}),
\lb{RR1}
\en
where we again understand $\tq'_1=0$.
When $\kn$ is odd, we also set $\tq'_{\kn-1}=0$ in \rlb{RR1} if $j=k-2$ for the same reason as discussed in footnote~\ref{fn:Dmax}.
When $\kn$ is even, we have an exceptional situation with
\eq
\bD'_{\kn-1}=\AYX\kk(\bC'_{\kn-1})=\AXY_{(2\kn-2,0)\app(2\kn-3)}(\bE'_{\kn-1}),
\en
which yields
\eq
\kappa\,q'-\tq'_{\kn-1}=0.
\lb{RR2}
\en

For an even $j$ such that $j\ne\kn-2$, we observe
\eq
\bD'_j=\AYZ\kk(\bC'_j)=\AYX_{(j+\kn-1,0)\app(j+\kn-2,0)}(\bE'_j)=\AYX_{(j,0)\app(j+1,0)}(\bE'_{j+1}),
\lb{D'gen3}
\en
which, with \rlb{qC'j}, implies
\eq
\kappa\,q'-\tq'_j-\tq'_{j+1}=0\quad(\text{even $j$ s.t. $j\ne\kn-2$}).
\lb{RR3}
\en
Finally, when $\kn$ is even, we have
\eq
\bD'_{\kn-2}=\AYZ\kk(\bC'_{\kn-2})=\AYX_{(2\kn-3,0)\app(2\kn-4,0)}(\bE'_{\kn-2})=\AYZ_{(\kn-2,0)\app(\kn-1,0)}(\bE'_{\kn-1}),
\lb{D'gen2}
\en
 which leads to
 \eq
 \kappa\,q'-\tq'_{\kn-2}+\tq'_{\kn-1}=0.
 \lb{RR4}
 \en
 
 When $\kn$ is odd, we sum up \rlb{RR1} and \rlb{RR3} from $j=1$ to $\kn-2$ to get
 \eq
 (\kn-2)\kappa\,q'+\tq'_1+\tq'_{\kn-1}=0,
  \lb{cond3}
 \en
 which implies $q'=0$ because $\tq'_1=\tq'_{\kn-1}=0$ as we discussed.
 When $\kn$ is even, we sum up \rlb{RR1} and \rlb{RR3} from $j=1$ to $\kn-3$ to get
  \eq
 (\kn-3)\kappa\,q'+\tq'_1+\tq'_{\kn-2}=0.
 \en
 We also find from \rlb{RR2} and \rlb{RR4} that $\tq'_{\kn-2}=2\kappa\,q'$, and hence
 \eq
 (\kn-1)\kappa\,q'=0.
 \lb{cond4}
 \en
 This concludes our proof that $q'=0$.
 
 \medskip\noindent
 {\bf Remark:}~The final conditions for showing $q=q'=0$ in the previous and the present subsections are either of the form $(\kn-2)\kappa\,q=0$, $(\kn-2)\kappa\,q'=0$ as in \rlb{cond1}, \rlb{cond3} or $(\kn-1)\kappa\,q=0$, $(\kn-1)\kappa\,q'=0$ as in \rlb{k-1kappa}, \rlb{cond4}.
It is worthwhile to note that the latter corresponds to cases in which the relevant prouct is written in the doubling-product form \rlb{DB1} or \rlb{DB2}.

 \subsection{Conserved quantities with $\km=2$}\label{S:k=2}
 Let us prove Thorem~\ref{T:k=2}, which determines the two-body part of a local conserved quantity $\hQ$ with $\km=2$.
 We first note that Lemma~\ref{L:manyshifts} in Section~\ref{S:reduction} with $\km=2$ implies the following.
  \begin{lemma}\label{L:k=2}
 If $\hQ$ is a local conserved quantity of the form \rlb{Q} with $\km=2$, then the coefficient $\qA$ is nonzero only when $\supp\bA=\{u,v\}$ or $\supp\bA=\{u\}$ with $u,v\in\La$ and $|u-v|=1$.
 \end{lemma}
 \noindent{\em Proof:}\/
 Lemma~\ref{L:manyshifts} shows any $\bA\in\PL$ such that $\wido\bA=2$ and $\qA\ne0$ must have $\wid_\alpha\bA=1$ for all $\alpha=2,\ldots,d$.
Noting that the same statement holds for other directions than the 1-direction, we get the desired statement.~\qedm

\medskip
We shall prove Theorem~\ref{T:k=2} by applying techniques that were developed in the foregoing subsections.
We shall be brief about details.

\begin{figure}
\centerline{\epsfig{file=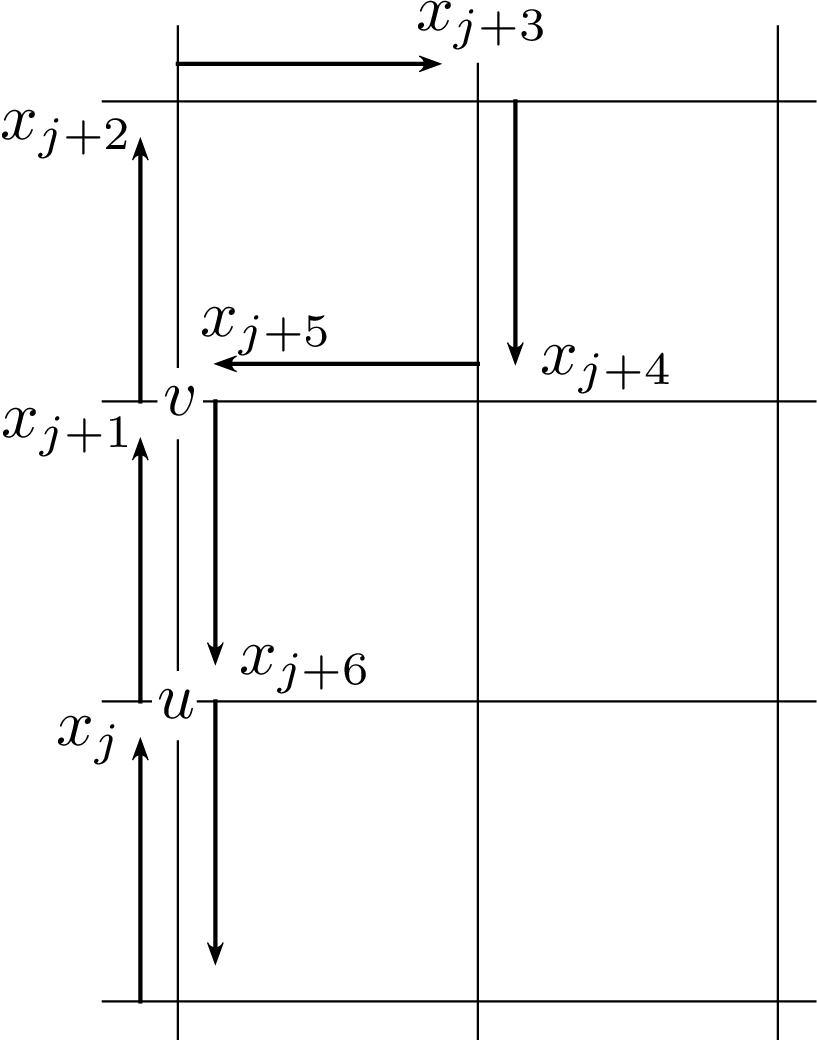,width=4.5truecm}}
\caption[dummy]{
The figure represents a small portion of the square lattice.
Fix $u,v\in\La$ such that $|u-v|=1$.
It is always possible to construct a locally-self-avoiding walk in which $x_j=u$ and $x_{j+1}=v$ for some $j$.
Then, the construction in the figure gives $x_{j+5}=v$ and $x_{j+6}=u$.
Clearly, $j$ and $j+5$ have different parity. 
}
\label{f:LSAW}
\end{figure}

We first show the terms $\hX_u\hX_v$, $\hY_u\hY_v$, and $\hZ_u\hZ_v$ have the desired coefficients as in \rlb{Q=H}.
Take a sequence $(x_1\ldots,x_N)$ with $x_j\in\La$ such that $|x_j-x_{j+1}|=1$ for all $j=1,\ldots,N-1$, and $x_j\ne x_{j+2}$ for all $j=1,\ldots,N-2$.
One may call such a sequence a locally-self-avoiding walk.
Observe, for any odd $j$, that
\eq
\hX_{x_j}\hZ_{x_{j+1}}\hY_{x_{j+2}}=\AYX_{x_{j+2}\app x_{j+1}}(\hX_{x_j}\hX_{x_{j+1}})=\AXY_{x_j\app x_{j+1}}(\hY_{x_{j+1}}\hY_{x_{j+2}}),
\en
which leads to
\eq
q_{\hY_{x_{j+1}}\hY_{x_{j+2}}}=\kappa\,q_{\hX_{x_j}\hX_{x_{j+1}}}\quad(\text{odd $j$}).
\lb{qYYqXX}
\en
One similarly finds
\eq
q_{\hX_{x_{j+1}}\hX_{x_{j+2}}}=\kappa^{-1}\,q_{\hY_{x_j}\hY_{x_{j+1}}}\quad(\text{even $j$}).
\lb{qYYqXX2}
\en
Now, write $q_{\hX_1\hX_2}$ as $\eta\Jx$ with some $\eta\in\bbC$.
Then, one finds from \rlb{qYYqXX} and \rlb{qYYqXX2} that
\eq
q_{\hX_{x_j}\hX_{x_{j+1}}}=\eta\,\Jx\ (\text{odd $j$}),\quad 
q_{\hY_{x_j}\hY_{x_{j+1}}}=\kappa\eta\,\Jx=\eta\,\Jy\ (\text{even $j$}).
\lb{qXXqYY}
\en
Note that it is possible to construct a locally-self-avoiding walk  $(x_1\ldots,x_N)$ that visits every neighboring pair $\{u,v\}$ at least twice with different parity, i.e., there are odd $n$ and even $m$ such that $\{u,v\}=\{x_n,x_{n+1}\}=\{x_m,x_{m+1}\}$.
This fact can be proved in a constructive manner as is described in Figure~\ref{f:LSAW} and its caption.
Then, \rlb{qXXqYY} implies
\eq
q_{\hX_u\hX_v}=\eta\,\Jx,\quad q_{\hY_u\hY_v}=\eta\,\Jy,
\en
for any $u,v\in\La$ with $|u-v|=1$.
The same argument shows $q_{\hZ_u\hZ_v}=\eta\,\Jz$, both for nonzero and zero $\Jz$.

It remains to show that the products of different Pauli matrices, e.g., $\hX_u\hY_v$, are prohibited.
The argument is by now standard.
Consider the products
\eq
\bA_1=\hX_{(1,0)}\hY_{(2,0)},\quad
\bA_2=\hY_{(2,0)}\hX_{(3,0)},\quad
\bA_3=\hX_{(2,1)}\hY_{(2,0)},
\en
and
\eq
\bB_1=\hX_{(1,0)}\hZ_{(2,0)}\hX_{(3,0)},\quad
\bB_2=\hX_{(1,0)}\hZ_{(2,0)}\hX_{(2,1)},\quad
\bB_3=\hX_{(2,1)}\hZ_{(2,0)}\hX_{(3,0)}.
\en
Observing that
\eqg
\bB_1=\AXY_{(3,0)\app(2,0)}(\bA_1)=\AXY_{(1,0)\app(2,0)}(\bA_2),\\
\bB_2=\AXY_{(2,1)\app(2,0)}(\bA_1)=\AXY_{(1,0)\app(2,0)}(\bA_3),\\
\bB_3=\AXY_{(2,1)\app(2,0)}(\bA_2)=\AXY_{(3,0)\app(2,0)}(\bA_3),
\eng
we find
\eq
q_{\bA_1}+q_{\bA_2}=0,\quad
q_{\bA_1}+q_{\bA_3}=0,\quad
q_{\bA_2}+q_{\bA_3}=0,
\en
which shows $q_{\bA_1}=q_{\bA_2}=q_{\bA_3}=0$.
Other cases can be treated in the same manner.

\section{Discussion}\label{S:discussion}
We have proved that the $S=\frac{1}{2}$ quantum spin model on the $d$-dimensional hypercubic lattice with the Hamiltonian \rlb{H} does not have nontrivial local conserved quantities under the condition that $\Jx\ne0$ and $\Jy\ne0$ if $d\ge2$.
That we are in two or higher dimensions is essential since there are one-dimensional models with $\Jx\ne0$ and $\Jy\ne0$, i.e., the XY or the XYZ model, that possess nontrivial local conserved quantities.

Although our proof is tailored to the XY and XYZ models on the hypercubic lattice, it is based on some general strategies.
In particular, the first part of the proof, in which we reduce the higher-dimensional problem to a one-dimensional problem, is rather general.
It is clear that our result can be extended to a larger class of models in high dimensions.

Let us start with an obvious extension.
The most general Hamlitonian for the $S=\frac{1}{2}$ system with uniform nearest neighbor interactions read
\eqa
 \hH_\mathrm{gen}=&-\sum_{u\in\La}\sum_{\alpha=1}^d\bigl\{\Jxx\,\hX_u\hX_{u+\bea}+\Jyy\,\hY_u\hY_{u+\bea}+\Jzz\,\hZ_u\hZ_{u+\bea}
\nl&\hspace{43pt}+\Jxy\,\hX_u\hY_{u+\bea}+\Jyx\,\hY_u\hX_{u+\bea}
+\Jyz\,\hY_u\hZ_{u+\bea}\nl&\hspace{43pt}+\Jzy\,\hZ_u\hY_{u+\bea}
+\Jzx\,\hZ_u\hX_{u+\bea}+\Jxz\,\hX_u\hZ_{u+\bea} 
 \bigr\}
 \nl&-\sum_{u\in\La}\bigl\{\hx\,\hX_u+\hy\,\hY_u+\hz\,\hZ_u\bigr\},
 \lb{Hgen}
 \ena
 where $\bea$ is the unit vector in the $\alpha$-direction.
 When the interactions have inversion symmetry 
 \eq
 \Jxy=\Jyx,\quad\Jyz=\Jzy,\quad\Jzx=\Jxz,
 \lb{inv}
 \en
 the matrix
 \eq
 \sfJ=\pmat{\Jxx&\Jxy&\Jxz\\\Jyx&\Jyy&\Jyz\\\Jzx&\Jzy&\Jzz},
 \en
is real symmetric.
By diagonalizing $\sfJ$ by an orthogonal matrix, the Hamiltonian can be transformed into the original form \rlb{H}, as is observed in \cite{YamaguchiChibaShiraishi2024a,YamaguchiChibaShiraishi2024b}.
If the rank of $\sfJ$ is two or three, then the transformed model satisfies the condition $\Jx\ne0$ and $\Jy\ne0$ necessary for our proof.
Our theorems are valid as they are in this case.
If the rank of $\sfJ$ is one, then the transformed Hamiltonian is that of the quantum Ising model with a magnetic field.
Our theorems do not cover these models, but we learned that Chiba had controlled them \cite{Chiba2024b}.

\begin{figure}
\centerline{\epsfig{file=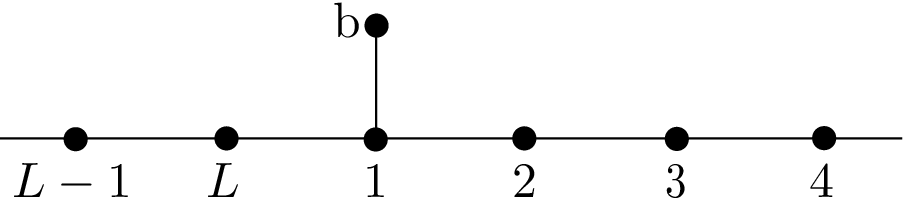,width=7truecm}}
\caption[dummy]{
The simplest geometry in which our proof works.
We can extend our proof to, e.g., the Hamiltonian $\hH=-\sum_{x=1}^L\{\hX_x\hX_{x+1}+\hY_x\hY_{x+1}\}-(\hX_1\hX_\mathrm{b}+\hY_1\hY_\mathrm{b})$, where we denoted sites on the periodic chain as $x=1,\ldots,L$ and the extra site in the branch as b.
For the proof, one only needs to identify the site b with $(\kn-1,1)$ in sections~\ref{S:k=3}, \ref{S:CYX}, and \ref{S:CXX}.
}
\label{f:branch}
\end{figure}

We have studied the simple Hamiltonian \rlb{H} with uniform and (spatially) isotropic interactions, but these conditions are not mandatory.
One may treat various models where the exchange interactions depend nontrivially on the coordinate directions with suitable extensions of our method.
It is worth mentioning that our proof of the absence of nontrivial local conserved quantities works for the same class of models defined on a ladder, namely, the two-dimensional $L\times 2$ lattice.
In fact, it even works for models on the periodic chain with one extra branch sticking out of it.
See Figure~\ref{f:branch}.
This means that the ``high-dimensional structure'' is already present in these simple geometries.
We also believe that one can prove the absence of nontrivial local conserved quantities in some interacting models of fermions in two or higher dimensions.
The Hubbard model is a suitable candidate in which one may try combining the analysis in \cite{Fukai2023} with the method developed here.

The current proof for models in two or higher dimensions is technically simpler than the corresponding proofs in one-dimensional models \cite{Shiraishi2019,Chiba2024a,ParkLee2024a,Shiraishi2024,ParkLee2024b,HokkyoYamaguchiChiba2024,YamaguchiChibaShiraishi2024a,YamaguchiChibaShiraishi2024b,Shiraishi2025}.
The simplicity reflects the natural tendency that the models are ``less integrable'' in dimensions two or higher.
The fact that the absence of nontrivial local conserved quantities can be proved even for the simplest XX model (obtained by setting $\Jx=\Jy\ne0$ and $\Jz=\hx=\hy=\hz=0$ in \rlb{H}) is a clear indication of this tendency.
We nevertheless admit that the latter half of the proof described in Sections~\ref{S:CYX} and \ref{S:CXX} still relies on craftsmanship.
It requires an adequate choice of a sequence of products and a careful (and tedious) treatment of the resulting conditions.
It is highly desirable to develop a general scheme that enables us to treat the intricate network of conditions, $\cB=0$ for all $\bB\in\PL$, in a more systematic manner.
Such a novel scheme not only simplifies the proof but is also expected to make the theory applicable to a larger class of models.\footnote{
The recent work by Hokkyo \cite{Hokkyo2025} is clearly a nontrivial first step.
}

\medskip
We believe that the proof of the absence of nontrivial local conserved quantities for standard quantum models is interesting and meaningful by itself.
It gives a strong indication that the models are not ``solvable'' and sheds light on the fundamental structure of the operator algebra of quantum many-body systems.
Having said that, we also believe it is desirable to have further results that are directly related to the long-time dynamics of the system, namely, the presence or the lack of thermalization or equilibration.

A crucial open question in this direction is to develop a similar non-existence theorem for quasi-local conserved quantities.
Roughly speaking, a quasi-local conserved quantity is an infinite sum (with suitably decaying coefficients) of products of Pauli matrices that commutes with the Hamiltonian.
See Appendix~\ref{S:QuasiLocal} for a more careful specification.
In the one-dimensional $S=\frac{1}{2}$ XXZ model, which is a well-known integrable model, it has been shown that quasi-local conserved quantities exist and play an essential role in determining the equilibrium state obtained by the long-time unitary evolution \cite{Prosen2014,IlievskiDeNardisWoutersCauxEsslerProsen2015}.

Recall that the current method of the proof of the absence of nontrivial local conserved quantities relies crucially on the existence of the maximum width $\km$.
It is likely that an essentially novel idea is necessary to develop a theory that controls quasi-local conserved quantities.
Developing the above-mentioned systematic scheme for treating the intricate network of conditions may be one possible direction.
In Appendix~\ref{S:QuasiLocal}, we give a very preliminary discussion about the absence of quasi-local conserved quantities.

Another important direction is to relate the current line of studies to that of the operator growth or the Krylov complexity \cite{ParkerCaoAvdoshkinScaffidiAltman2019,Cao2021,MuckYang2022,NandyMatsoukasRoubeasMartinezAzconaDymarskyCampo2024}.
It seems apparent that the lack of nontrivial local conserved quantities has some nontrivial implications on the nature of the operator growth.
Roughly speaking, it implies that a sufficiently complex operator must change in time.
It is a challenging problem to make the connection quantitative and, ultimately, rigorously establish the relation between the absence of conserved quantities and operator growth.
As a first step, we prove lower bounds for the Lanczos coefficients characterizing operator growth in Appendix~\ref{S:OG}.
We, in particular, establish that the general class of models treated in the present paper exhibit clear signature of quantum chaos.
We also address the closely related issue of imaginary-time evolution of operators in Appendix~\ref{s:ITE}.

Last but not least, we stress that the present approach initiated in \cite{Shiraishi2019} deals solely with the operator algebra, i.e., commutation relations between the Pauli matrices, and does not deal with quantum states.
This restriction certainly has a strategic advantage, which enabled the proof of the absence of nontrivial local conserved quantities.
Nevertheless, it goes without saying that a rigorous theory that characterizes a non-integrable system in terms of its quantum states is in strong demand.
We should explore ways to reflect the nature of states or subspaces of states in a similar approach.
Although we still cannot imagine what kind of theory is possible, one of the ultimate goals may be to provide a rigorous basis for ETH \cite{Deutsch1991,Srednicki1994,RigolSrednicki2012,DAlessioKafriPolkovnikovRigol2016}.
Such a project will be highly nontrivial, at least because the present approach to concentrate on the absence of nontrivial local conserved quantities does not distinguish between non-integrable models that satisfy ETH and those that do not because of quantum many-body scars (see footnote~\ref{fn:scar}).\footnote{
For example, the PXP model \cite{Bernienetal2017,TurnerMichailidisAbaninSerbynPapic2018}, the Majumdar-Ghosh model \cite{MoudgalyaRachelBernevigRegnault2018}, and the Affleck-Kennedy-Lieb-Tasaki model \cite{MoudgalyaRegnaultBernevig2018,MarkLinMotrunich2020} have scars (and hence fail to satisfy ETH), while these models are proved to possess no nontrivial local conserved quantities in \cite{ParkLee2024a}, \cite{Shiraishi2024}, and \cite{ParkLee2024b,HokkyoYamaguchiChiba2024}, respectively.
}

\appendix
\section{Appendix: Three related issues}
\subsection{Quasi-local conserved quantities}\label{S:QuasiLocal}
In this appendix, we shall give some preliminary discussion concerning quasi-local conserved quantities.
We first present an elementary and natural definition of a quasi-local conserved quantity in the setting of infinite systems.
It may be useful to give a clear definition that is free from subtle finite-size effects.
We then show that an extremely preliminary no-go theorem can be proved by simply modifying the proof for the absence of nontrivial local conserved quantities in the main text.

We consider a quantum spin system on the $d$-dimensional hypercubic lattice $\Zd$.
We assume $S=\frac{1}{2}$ for notational simplicity, but the extension to higher spins is automatic.\footnote{
One only needs to identify and fix a set of operators (for a single spin) that span the space of whole operators.
Recall that the set $\{\hS^{\rm x},\hS^{\rm y}, \hS^{\rm z}, \hI\}$ is insufficient for $S\ge1$.
}

As in \rlb{A}, we define a product of Pauli matrices as 
\eq
\bA=\bigotimes_{u\in S}\hA_u,
\lb{bAZd}
\en
where $S=\sup\bA$ is a noneqmpty finite subset of $\Zd$ and $\hA_u\in\{\hX_u,\hY_u,\hZ_u\}$.
The width $\wid\bA$ is defined as in Section~\ref{S:main}.
We denote by $\PZd$ the set of all products.
We write our Hamiltonian formally as\footnote{
We say this is formal because we do not try to make any sense of the infinite sum.
One may regard \rlb{HZd} as giving a list of well-defined operators $\hh_u$ for every $u\in\Zd$.
}
\eq
\hH=\sum_{u\in\Zd}\hh_u,
\lb{HZd}
\en
where each $\hh_u$ is a self-adjoint operator that acts nontrivially only on a finite support $\supp\hh_u\subset\Zd$ with $\supp\hh_u\ni u$.
It is reasonable to assume translational invariance of $\hh_u$, but let us be general here and only assume there are constants $h_0$ and $w_0$ such that $\snorm{\hh_u}\le h_0$ and $\wid\supp\hh_u\le w_0$ for any $u\in\Zd$.

It is crucial to note that, although the Hamiltonian $\hH$ is only formally defined by \rlb{HZd}, the commutator between $\hH$ and a product $\bA$ can be defined as
\eq
[\hH,\bA]=\sumtwo{u\in\Zd}{(\supp\hh_u\cap\supp\bA\ne\emptyset)}[\hh_u,\bA].
\lb{HAZd}
\en
Since the right-hand side is an operator acting nontrivially only on a finite support, we can uniquely expand it as
\eq
[\hH,\bA]=\sum_{\bB\in\PZd}\la_{\bA,\bB}\,\bB.
\lb{HAZd2}
\en
Note that, for a given $\bB$, the coefficient $\la_{\bA,\bB}$ is nonzero only for a finite number of $\bA$.

We write the candidate of a quasi-local conserved quantity as a formal sum
\eq
\hQ=\sum_{\bA\in\PZd}\qA\,\bA,
\lb{QZd}
\en
where $\qA\in\bbC$.\footnote{\label{fn:localforZd}
One has the candidate of a local conserved quantity by restricting the summation to those $\bA$ with $\wid\bA\le\km$.
}
For the quantity to be meaningful, we need to assume a certain decay property of the coefficients $\qA$.
A reasonable choice is
\eq
|\qA|\le F(\wid\bA),
\lb{qAF}
\en
where $F(k)$ is a positive decreasing function of $k\in\bbN$ that decays sufficiently fast to zero as $k\up\infty$.
The known examples of quasi-local conserved quantities in the one-dimensional XXZ model have exponentially decaying coefficients \cite{Prosen2014}.
From \rlb{HAZd2} and \rlb{QZd}, we can rewrite (or, more precisely, define) the commutator of $\hH$ and $\hQ$ as
\eq
[\hH,\hQ]=\sum_{\bA\in\PZd}\qA\,[\hH,\bA]=\sum_{\bB\in\PZd}\cB\,\bB,
\lb{HQZd}
\en
where
\eq
\cB=\sum_{\bA\in\PZd}\la_{\bA,\bB}\,\qA.
\lb{cBZd}
\en
It is crucial that \rlb{cBZd} defines $\cB$ in terms of a finite sum, while \rlb{HQZd} is a formal sum.

We can then define the notion of a quasi-local conserved quantity.
\begin{definition}
We say that nonzero $\hQ$ of the form \rlb{QZd} satisfying the decay property\footnote{
To be rigorous, our definition is incomplete since we have not specified the decay function $F(k)$.
But, for the moment, we want to be flexible about the details.
} \rlb{qAF} is a quasi-local conserved quantity if and only if it is not a local conserved quantity and it holds that $\cB=0$ for any $\bB\in\PZd$.
\end{definition}

Our goal is to prove a no-go theorem that is parallel to our main theorem, Theorem~\ref{T:main}, or theorems proved in the series of papers \cite{Shiraishi2019,Chiba2024a,ParkLee2024a,Shiraishi2024,ParkLee2024b,HokkyoYamaguchiChiba2024,YamaguchiChibaShiraishi2024a,YamaguchiChibaShiraishi2024b,Shiraishi2025}, and rules out the existence of quasi-local conserved quantities with a suitable decay function $F(w)$.
For the moment, this is a grand challenge and such a proof seems to require an essentially new idea.

In order to demonstrate that such a statement is not impossible, we shall describe an extremely preliminary observation that follows from a straightforward modification of the proof in the main body of the present paper.
For simplicity, we consider the $d$-dimensional XX model\footnote{
There are no essential obstacles in proving a similar result for the model with full XYZ interaction and magnetic fields.
We simply wanted to avoid treating the ratios of the parameters (like $\kappa=\Jy/\Jx$) that appear in various relations when treating the general models.
} on the infinite lattice $\Zd$ with the local Hamiltonian
\eq
\hh_u=-\sum_{\alpha=1}^d\bigl\{\hX_u\hX_{u+\bea}+\hY_u\hY_{u+\bea}\bigr\},
\en
where $\bea$ is the unit vector in the $\alpha$-direction.
This lcoal Hamiltonian gives \rlb{H} with $\Jx=\Jy=1$, $\Jz=\hx=\hy=\hz=0$ (with $\La$ replaced by $\Zd$) when summed over $u\in\Zd$.
We can then exclude the existence of a quasi-local conserved quantity that exhibits extremely quick decay.
\begin{proposition}
There exist no quasi-local conserved quantities with $F(k)$ satisfying
\eq
\lim_{k\up\infty}(4d^2)^k(k!)^{d+3}F(k)=0.
\lb{Fcrazydecay}
\en
\end{proposition}
We should repeat that this is only meant to be a demonstration that a no-go theorem is at least possible.
We do not have any physical justification for the assumption \rlb{Fcrazydecay} about the super-quick decay.
The challenge is to find a new argument that enables us to relax this unphysical assumption and rule out quasi-local conserved quantities with, say, exponentially decaying $F(k)$.

\medskip
\noindent
{\em Proof:}\/
As we mentioned above, the following proof is a straightforward modification of that in the main text.
We also note that the following bounds may be improved (but not in an essential manner) since we did not make any efforts to optimize constants in inequalities.

Suppose that there is a quasi-local conserved quantity $\hQ$.
Take and fix $k_0$ such that there is $\bA\in\PZd$ with $\wid\bA=k_0$ and $\qA\ne0$.
Note that such $k_0$ exists if $\hQ$ exists.

We take $k>k_0$, and try to repeat the proof in the main text by replacing $\km$ by $k$.
Let $\bA\in\PZd$ be such that $\wid\bA=k$.  Without loss of generality, we may assume $\wido\bA=k$.
Note that the number of sites in $\supp\bA$ is bounded by $k^d$.

Let us start from the proof in Section~\ref{S:reduction}.
There, we repeatedly used the condition $\cB=0$ for $\bB$ such that $\wido\bB=\km+1$, along with the fact that one inevitably has $q_{\bA'}=0$ for any $\bA'$ with $\wido\bA'>\km$.
In the present setting, we still have $\cB=0$, but the corresponding coefficients $q_{\bA'}$ may not be zero anymore.
This means that whenever we use the condition $\cB=0$, we may get some error coming from those $\bA'$ that were not counted in Section~\ref{S:reduction}.
Since there are at most $2dk^d$ such $\bA'$'s and the condition $\cB=0$ is used at most $k+1$ times, we see that the identity $\qA=0$ in Lemmas~\ref{L:inheritance} and \ref{L:manyshifts} should be replaced by
\eq
|\qA|\le 2dk^d(k+1)\,\newQ(k+1),
\lb{qAsup}
\en
were we defined
\eq
\newQ(k)=\suptwo{\bA'\in\PZd}{(\wido\bA'\ge k)}|q_{\bA'}|.
\lb{Q(k)}
\en

We next move to the proof in Sections~\ref{S:CYX} and \ref{S:CXX}.
There, we used the conditions $c_{\bD_j}=0$ or $c_{\bD'_j}=0$ repeatedly for $\km-1$ or $\km-2$ times.
We could neglect the contributions from products with  $\wido=\km$ and $\wid_2=2$ because of Lemma~\ref{L:manyshifts}.
We now should assume that these products may have nonzero contributions bounded as \rlb{qAsup}.
Since there are at most $2d(k+1)$ such products, the conclusions of these sections should be modified as
\eq
|q_{\bC_{\rm YX}}|\le 2d(k+1)(k-1) \times (\text{right-hand side of \rlb{qAsup}})
+3\,\newQ(k+1),
\lb{qCsup}
\en
where the second term, which is indeed minor, counts possible contributions from products with $\wido=k+1$.
We have the same bound for $q_{\bC_{\rm XX}}$.

From \rlb{qAsup} and \rlb{qCsup}, we conclude that
\eq
|\qA|\le 4d^2(k+1)^{d+3}\,\newQ(k+1),
\en
for any $\bA\in\PZd$ such that $\wido\bA=k$.
Since this implies $\newQ(k)\le 4d^2(k+1)^{d+3}\,\newQ(k+1)$, we find by iteration that
\eq
\newQ(k_0)\le(4d^2)^{k-k_0}\biggl(\frac{k!}{k_0!}\biggr)^{d+3}\newQ(k)\le(4d^2)^{k-k_0}\biggl(\frac{k!}{k_0!}\biggr)^{d+3}F(k),
\en
where we noted the assumption \rlb{qAF} and the definition \rlb{Q(k)} imply $\newQ(k)\le F(k)$.
By letting $k\up\infty$, we see that the assumed quick decay \rlb{Fcrazydecay} implies $\newQ(k_0)=0$.
But this contradicts the existence of $k_0$.~\qedm

\subsection{Spectrum generating algebra}\label{S:SGA}
In a general quantum system with Hamiltonian $\hH$, an operator $\hQ$ satisfying
\eq
[\hH,\hQ]=\mu\,\hQ,
\lb{HQQ}
\en
with some $\mu\in\bbR\backslash\{0\}$ is said to be an element of a spectrum generating algebra (SGA).
Note that $\hH\ket{\Phi}=E\ket{\Phi}$ and \rlb{HQQ} imply $\hH\hQ^n\ket{\Phi}=(E+n\mu)\hQ^n\ket{\Phi}$ for $n=1,2,\ldots$.
Thus $\hQ^n\ket{\Phi}$ is an energy eigenstate if nonvanishing.
We see that the existence of $\hQ$ satisfying \rlb{HQQ} implies the model has towers of energy eigenstates.
See, e.g., \cite{MarkLinMotrunich2020,MoudgalyaRegnaultBernevig2020} and references therein.\footnote{
These references mainly concern restricted spectrum generating algebras (RSGA), which consist of operators satisfying \rlb{HQQ} on certain subspaces.
}
A standard example is the spin-raising and lowering operators $\hSt^\pm$ in the Heisenberg model under a magnetic field in the z-direction, namely, the model \rlb{H} with $\Jx=\Jy=\Jz$, $\hx=\hy=0$, and $\hz\ne0$.
One readily sees that 
\eq
\hSt^\pm=\frac{1}{2}\sum_{u\in\La}(\hX_u\pm\ri\hY_u),
\lb{Qpm}
\en
satisfy
\eq
[\hH,\hSt^\pm]=\mp2\hz\,\hSt^\pm.
\en

Interestingly, we can provide an almost complete characterization of SGA by a slight modification of the proof in the main body of the paper.
In exactly the same setting, we can prove that SGA, if exists, consists of sums of one-body operators.
\begin{proposition}
Under the same conditions as in Theorem~\ref{T:main}, any operator $\hQ$ of the form \rlb{Q} with $2\le\km\le L/2$ does not satisfy \rlb{HQQ}.
\end{proposition}

\noindent{\em Proof:}\/
From \rlb{Q} and \rlb{HQexp}, we see that \rlb{HQQ} is equivalent to the condition
\eq
\cB=\mu\,\qB,
\lb{cB=qB}
\en
for all $\bB\in\PL$.
This corresponds to the condition \rlb{c=0} for a conserved quantity.

Note first that all the results in Section~\ref{S:reduction} is based on the condition \rlb{c=0}, $\cB=0$, for $\bB$ such that $\wido\bB=\km+1$.
In the present setting, we have $\qB=0$ for such $\bB$ by definition, and hence the new condition \rlb{cB=qB} is equivalent to \rlb{c=0}.
This means all the results in Section~\ref{S:reduction} are valid as they are.

Let $\km\ge3$.
We recall that all the results in Sections~\ref{S:preliminaries}, \ref{S:CYX}, and \ref{S:CXX} are based on the condition $c_{\bD_j}=0$.
But, since $\wido\bD_j=\km$ and $\wid_2\bD_j=2$, Lemma~\ref{L:manyshifts} shows $q_{\bD_j}=0$.
Then, again the new condition $c_{\bD_j}=\mu\,q_{\bD_j}$ reduces to $c_{\bD_j}=0$.
We conclude all the results in Sections~\ref{S:preliminaries}, \ref{S:CYX}, and \ref{S:CXX} are valid as they are.
We see that $\hQ$ satisfying \rlb{HQQ} does not exist.

For $\km=2$, we find from a straightforward extension of the proof of Theorem~\ref{T:k=2} that $\hQ$ is written as $\hQ=\eta\hH+\hQ^{(1)}$, exactly as in \rlb{Q=H}.
But this does not satisfy \rlb{HQQ} since $[\hH,\hQ]=0$.~\qedm

\subsection{Signature of quantum chaos in the Lanczos coeffecients}\label{S:OG}
The universal operator growth hypothesis of Parker, Cao, Avdoshkin, Scaffidi, and Altman \cite{ParkerCaoAvdoshkinScaffidiAltman2019} states that the growth of the Lanczos coefficients provides an essential characterization of the dynamics of a quantum many-body system.
It is believed that the Lanczos coefficients grow linearly in the recursion number $n$ in a chaotic system.
See also \cite{Cao2021,MuckYang2022,NandyMatsoukasRoubeasMartinezAzconaDymarskyCampo2024} and a large number of references therein.
In this Appendix, we prove new lower bounds for the growth of the Lanczos coefficients.\footnote{
A YouTube video discussing the results in Appendices~\ref{S:OG} and \ref{s:ITE} and some additional results is available \cite{video3}.} 

As discussed in Section~IV.B of the seminal paper \cite{ParkerCaoAvdoshkinScaffidiAltman2019}, the earlier work by Bouch \cite{Bouch2015} on the singularity in the complex-time evolution in a particular two-dimensional quantum spin system indicates that the Lanczos coefficients $b_n$ of the same model grow linearly in $n$.
Cao \cite{Cao2021} proved the same result for a much more realistic model, namely, the quantum Ising model under a transverse magnetic field in two or higher dimensions.
Here, we prove essentially the same result on the growth of Lanczos coefficients in the general class of models treated in the present paper, thus partially confirming the conjectures in \cite{ParkerCaoAvdoshkinScaffidiAltman2019,Cao2021}.

Let us first state our main result.  See below for definitions and notations;  the Lanczos coefficients are defined in \rlb{bdef}.
Our theorem heavily relies on the work of Bouch \cite{Bouch2015}.  
\begin{theorem}[Linear growth of the Lanczos coefficients]\label{T:OG}
Consider a $S=\frac{1}{2}$ quantum spin system on the $d$-dimensional hypercubic lattice $\Zd$ with $d\ge2$.
We treat the general Hamiltonian $\hH$ on $\Zd$ of the form \rlb{H} with $\Jx\ne0$ and $\Jy\ne0$.
Then there exists a positive constant $\alpha$ and an infinite set $G$ of positive integers such that the Lanczos coefficients $b_n$ for the initial operator $\hX_o$ satisfy
\eq
\prod_{j=1}^nb_j\ge \alpha^n\,n!,
\lb{bn>Cn}
\en
for any $n\in G$.
\end{theorem}
Roughly speaking, the theorem states that the Lanczos coefficients grow with $n$ as $b_n\gtrsim \alpha\,n$, provided that $b_n$ depends on $n$ almost monotonically.\footnote{This is not the case in general, but we expect that $b_n$ in the present models behave nicely.}
We clearly see a signature of quantum chaos in the general models in two or higher dimensions.
One can prove the same lower bound for the Ising model under a transverse magnetic field by using techniques developed in \cite{Cao2021}.
If we consider the same model on the Bethe lattice, then the bound \rlb{bn>Cn} can be (easily) proved for any positive integer $n$.

\medskip
Let us give precise definitions, following \cite{ParkerCaoAvdoshkinScaffidiAltman2019,MuckYang2022,NandyMatsoukasRoubeasMartinezAzconaDymarskyCampo2024}.
As in Appendix~\ref{S:QuasiLocal}, we study a $S=1/2$ quantum spin system on the infinite lattice $\Zd$.
By a local operator, we mean a linear combination of a finite number of products of Pauli matrices as in \rlb{bAZd}.
For local operators $\hA$ and $\hB$, we define their inner product by $\sbkt{\hA,\hB}=\rho_\infty(\hA^\dagger\hB)$.
Here, $\rho_\infty$ denotes (the expectation value in) the infinite temperature Gibbs state defined as
\eq
\rho_\infty(\hA)=\frac{\Tr_{\calH_S}[\hA]}{\Tr_{\calH_S}[\hI]},
\en
where $S\subset\Zd$ is a sufficiently large subset that includes the support of $\hA$ and $\calH_S$ the corresponding Hilbert space.
This inner product is the Hilbert-Schmidt inner product with extra normalization.
It is worth noting that products of Pauli matrices \rlb{bAZd} satisfy the orthonormality
\eq
\sbkt{\bA,\bB}=\delta_{\bA,\bB}\ \text{for any $\bA,\bB\in\PZd$}.
\lb{<A,B>}
\en

Let $\hB_0$ be a local operator.
For simplicity, we assume $\hB_0$ is self-adjoint and $\sbkt{\hB_0,\hB_0}=1$.
In Theorem~\ref{T:OG}, we set $\hB_0=\hX_o$, where $o$ denotes the origin of $\Zd$.
We then define a series of self-adjoint local operators $\hB_0,\hB_1,\hB_2,\ldots$ recursively as
\eq
\hB_n=\ri\,[\hH,\hB_{n-1}],
\lb{BB}
\en
where $\hH$ denote the formal Hamiltonian on $\Zd$ as in \rlb{HZd}.
We here assume the local Hamiltonian $\hh_u$ corresponds to those in \rlb{H}.
Although the Hamiltonian $\hH$ itself is a formal object, the commutator with a local operator is well-defined as in \rlb{HAZd}.
The formal expression\footnote{See \rlb{Btrig} below for the precise definiton.} for the time-evolution of $\hB_0$
\eq
\hB(t)=e^{\ri\hH t}\,\hB_0\,e^{-\ri\hH t}=\sum_{n=0}^\infty\frac{t^n}{n!}\,\hB_n,
\lb{fB(t)}
\en
suggests that the series $\hB_0,\hB_1,\hB_2,\ldots$ contains information about the time-evolution of the operator $\hB_0$.

It is known that a convenient measure of the operator growth is given by the Gram-Schmidt-orthogonalized series $\hC_0,\hC_1,\hC_2,\ldots$ defined as follows.\footnote{
To be precise, we assume $\hB_0,\hB_1,\ldots,\hB_n$ are linearly independent for any $n=1,2,\ldots$.
This is valid for the case treated in Theorem~\ref{T:OG}.
}
We set $\hC_0=\hB_0$, and then define $\hC_n$ for $n=1,2,\ldots$ as a unique operator written as
\eq
\hC_n=\hB_n+\sum_{j=1}^{n-1}\alpha_j^{(n)}\hB_j,
\lb{CBdef}
\en
with some coefficients $\alpha_j^{(n)}$,
and satisfies
\eq
\sbkt{\hC_n,\hC_j}=0\ \text{for all $j=0,\ldots,n-1$}.
\en
Roughly speaking, $\hC_n$ is the part of $\hB_n$ that appears for the first time in the $n$-th recursion.
Then, for $n=1,2,\ldots$, the Lanczos coefficient for the series $\hC_0,\hC_1,\hC_2,\ldots$ is defined by
\eq
b_n=\sqrt{
\frac{\sbkt{\hC_{n},\hC_{n}}}{\sbkt{\hC_{n-1},\hC_{n-1}}}
},
\lb{bdef}
\en
or, equivalently, by
\eq
\prod_{j=1}^nb_j=\sqrt{\sbkt{\hC_{n},\hC_{n}}}.
\lb{bbbC}
\en

\medskip
\noindent{\bf Remark:}~It is known that the series $\hC_0,\hC_1,\hC_2,\ldots$ is determined by the recursion relation
\eq
\hC_n=\ri\,[\hH,\hC_{n-1}]+\frac{\sbkt{\hC_{n-1},\hC_{n-1}}}{\sbkt{\hC_{n-2},\hC_{n-2}}}\,\hC_{n-2},
\lb{monicrec}
\en
for $n=1,2,\ldots$, where we set $\hC_{-1}=0$.
See the discussion about the ``monic version" in \cite{MuckYang2022,NandyMatsoukasRoubeasMartinezAzconaDymarskyCampo2024}.
Our operators are related by $\hC_n=\ri^nO_n$ to those in \cite{MuckYang2022,NandyMatsoukasRoubeasMartinezAzconaDymarskyCampo2024}.
We do not use the recursion relation \rlb{monicrec} in what follows.

\medskip
\noindent{\em Proof of Theorem~\ref{T:OG}:}\/
We set $\hB_0=\hX_o$.
From the recursive definition \rlb{BB}, we see that $\hB_n$ is a superposition of products whose support contains at most $n+1$ sites.
Suppose that $\hB_n$ contains a term $a\bA$, where $\bA$ is a product whose support consists of $n+1$ sites, and $a\in\bbR$ is the associated coefficient.
Since the product $\bA$ cannnot appear in $\hB_j$ with $j=1,\ldots,n-1$, we see from \rlb{CBdef} that $\hC_n$ also contains the term $a\bA$.
From the orthonormality \rlb{<A,B>}, we see $\sbkt{\hC_n,\hC_n}\ge a^2$.
We thus get the desired \rlb{bn>Cn} from \rlb{bbbC} if we find an appropriate term $a\bA$ in $\hB_n$ such that $|a|\ge\alpha^n n!$.

\begin{figure}
\centerline{\epsfig{file=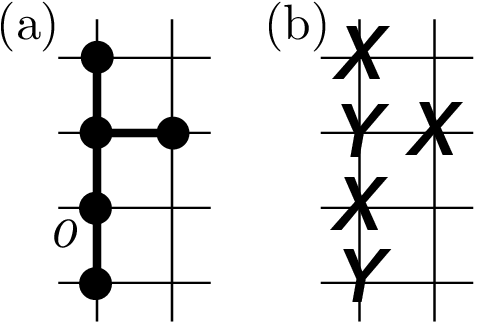,width=4truecm}}
\caption[dummy]{
(a) An example of a tree $(T,\tilde{T})$ with five sites (vertices) and four bonds (edges) on the square lattice.
The black dots represent sites in $T$, and the thick lines bonds in $\tilde{T}$.
(b)~The product $\bA$ with $\supp\bA=T$ generated by the specified procedure.
There are eight different ways to generate the product, i.e., $N(T)=8$.
}
\label{f:tree}
\end{figure}

We proceed in a constructive manner to find such a term.
We take a set $T\subset\Zd$ of $n+1$ sites, including the origin $o$, and denote by $\tilde{T}$ the collection of all bonds $\{u,v\}$ (i.e., a pair of sites with $|u-v|=1$) such that $u,v\in T$.
We assume $T$ is chosen so that the graph $(T,\tilde{T})$ is connected and contains no loops.
In other words, the graph $(T,\tilde{T})$ is a tree rooted at the origin $o$.
See Figure~\ref{f:tree}~(a).
We start from the initial operator $\hX_o$ and successively apply the appending operation $n$ times to generate a product whose support coincides with $T$.

Let us discuss the construction in more detail.
For $u\in T$, we denote by $d(u)$ the graph-theoretic distance between $o$ and $u$ on the graph $(T,\tilde{T})$.
Take $\{u,v\}\in \tilde{T}$ such that $d(v)=d(u)+1$, and suppose that we have a product in which $u$ is occupied by a Pauli matrix and $v$ is not.
We then apply the operation $\calA^{\hX\hX}_{v\app u}$ if $d(u)$ is odd and the operation $\calA^{\hY\hY}_{v\app u}$ if $d(u)$ is even to add a Pauli matrix to site $v$.
Repeating this process, we will end up with a product $\bA$ such that $\supp\bA=T$.
See Figure~\ref{f:tree}~(b).

It should be noted that, in general, there are multiple ways of (re)constructing the tree $(T,\tilde{T})$ according to the above procedure since the order of appending operations may be changed.
However, we always end up with the same $\bA$ with the same coefficients.
It is also crucial that reflecting the geometry of $T$, the product $\bA$ appearing in $\hB_n$ can be generated only by this procedure.
Therefore, if we denote by $N(T)$ the number of different ways to construct $(T,\tilde{T})$, we can bound the coefficient $a$ that comes with $\bA$ as
\eq
|a|\ge (2J_{\rm min})^n\,N(T),
\en 
where $J_{\rm min}=\min\{|\Jx|,|\Jy|\}$.
In \cite{Bouch2015}, Bouch proved, by devising an ingenious example, that there is a constant $C>1$ and an infinite set $G$ of positive integers such that for any $n\in G$, there exists a tree $(T,\tilde{T})$ with $n$ bonds rooted at the origin that can be constructed in at least $n!/C^n$ ways.
(See \cite{Tasaki2024} for an account of Bouch's proof.)
We thus find $|a|\ge\alpha^n\,n!$ with $\alpha =2J_{\rm min}/C$.

If we consider the Bethe lattice instead of $\Zd$, it is well known that, for any positive integer $n$, there exists a tree $(T,\tilde{T})$ with $n$ bonds that can be grown from the origin in at least $n!/C^n$ ways.\footnote{
Consider, for simplicity, the Bethe lattice with coordination number three.
The number of ways to grow a tree from the origin is $3\times4\times5\times\cdots\times(n+2)$.
The number of distinct trees with $n+1$ sites, on the other hand, is upper bounded by $3^{2n}$ since any tree with $n+1$ sites can be realized as a track of a random walk with $2n$ steps that starts and ends at $o$.
This means there exists a tree that can be grown in at least $3\times\ldots\times(n+2)/9^n>n!/9^n$ ways.
}
We thus have the desired lower bound \rlb{bn>Cn} for any positive $n$.~\qedm

\medskip
It is notable that Bouch's construction of trees that can be generated in ``too many'' ways essentially makes use of the fact that the lattice contains the full two-dimensional lattice $\bbZ^2$.
This is in contrast to the proof of our main theorem on the absence of nontrivial local conserved quantities, which works even on a quasi-one-dimensional lattice, including the ladder and the one in Figure~\ref{f:branch}.

The results by Bouch \cite{Bouch2015} and Cao \cite{Cao2021} with the above Theorem~\ref{T:OG} suggest that quantum chaos, or more precisely, the linear growth of the Lanczos coefficients $b_n$, is quite a robust phenomenon in dimensions two or higher.
Apparently, the same mechanism that makes use of the full two-dimensional plane is absent for one-dimensional models, about which we shall make some comments.

As a prototypical example, let us consider the $S=\frac{1}{2}$ XY chain with a magnetic field in the Y-direction, whose Hamiltonian is
\eq
\hH=-\sum_{j=1}^\infty\{\Jx\,\hX_j\hX_{j+1}+\Jy\,\hY_j\hY_{j+1}+h\,Y_j\},
\en
where we assume $\Jx\ne0$ and $\Jy\ne0$.
The model has been expected to be non-integrable if $h\ne0$.
In fact, it was proved by Yamaguchi, Chiba, and one us (N.S.) \cite{YamaguchiChibaShiraishi2024a,YamaguchiChibaShiraishi2024b} that the model with $h\ne0$ has no nontrivial local conserved quantities.
It is then expected from the universal operator growth hypothesis \cite{ParkerCaoAvdoshkinScaffidiAltman2019} that the Lanczos coefficients show linear growth with a logarithmic correction.
By using the simple strategy as in the proof of Theorem~\ref{T:OG}, we can prove the following weaker result.
\begin{theorem}\label{T:OG1d}
If $\Jx\ne0$, $\Jy\ne0$, and $h\ne0$, the Lanczos coefficients $b_n$ for the initial operator $\hX_0$ satisfy
\eq
\prod_{j=1}^nb_j\ge \gamma^n\,\lfloor\tfrac{n}{2}\rfloor!,
\lb{bn>1d}
\en
with a constant $\gamma>0$ for any $n=1,2,\ldots$.
\end{theorem}
The theorem roughly implies that the Lanczos coefficients $b_n$ grow at least as a constant times $\sqrt{n}$.\footnote{Interestingly, it was observed numerically that the Lanczos coefficients grow proportionally to $\sqrt{n}$ in interacting integrable models \cite{ParkerCaoAvdoshkinScaffidiAltman2019}.}
It is challenging to develop a more refined combinatorial estimate to prove the linear growth of $b_n$ in this model.

Notably, Cao \cite{Cao2021} proved that the moments $\mu_{2n}=(-1)^n\sbkt{\hB_0,\hB_{2n}}$ in the quantum Ising chain with the Hamiltonian $\hH=-\sum_{j=1}^L(\hZ_j\hZ_{j+1}+h\hX_j+h'\hZ_j)$ with $h\ne0$ and $h'\ne0$ satisfy
\eq
\mu_{2n}\gtrsim\Bigl(\frac{a\,n}{\log n}\Bigr)^{2n},
\en
with a constant $a>0$, where the symbol $\gtrsim$ indicates that the right-hand side shows the leading behavior when $n$ grows.
This implies that the Lanczos coefficients satisfy
\eq
\max_{j\in\{1,\ldots,n\}}b_j\gtrsim \frac{c\,n}{\log n},
\en
with a constant $c>0$.\footnote{
It holds in general that $\max_{j\in\{1,\ldots,n\}}b_j\ge(C_n\,\mu_{2n})^{1/(2n)}$, where $C_n=(2n)!/\{(n+1)!\,n!\}\simeq 4^n/\{\sqrt{\pi}\,n^{3/2}\}$ is the Catalan number.  See Appendix A2 of \cite{ParkerCaoAvdoshkinScaffidiAltman2019}.
}.
The linear growth with a logarithmic correction is precisely the expected behavior of Lanczos coefficients in one-dimensional chaotic systems.
We also note that Chiba proved that the same model has no local conserved quantities other than the Hamiltonian itself \cite{Chiba2024a}.

\medskip\noindent
{\em Proof of Theorem~\ref{T:OG1d}:}\/
We treat the case where $n$ is a multiple of four.  Other cases can be treated similarly with (minor) extra care.
Our strategy is again to locate a product that appears in $\hB_n$ but not in $\hB_j$ with $j<n$ and can be generated in multiple manners.
Starting from the initial operator $\hX_0$, we apply appending operations $n/2$ times to generate the product $\bigl(\bigotimes_{j=0}^{(n/2)-1}\hZ_j\bigr)\otimes\hX_{n/2}$.
Note that there is a unique way to do this.
We then use the magnetic filed term and take the commutator with $\hY_j$ with all $j=0,\ldots,(n/2)-1$ to generate $\bA=\bigotimes_{j=0}^{n/2}\hX_j$.
Since the commutator can be taken in an arbitrary order, there are $(n/2)!$ ways of performing the second step.
The same product $\bA$ may be generated by different processes, but an inspection shows it always comes with the same coefficient.\footnote{
In fact, $\bA$ is generated in $n!!$ ways, but this estimate does not essentially improve our conclusion.
}
We thus find that $\hC_n$ contains the term $a\bA$ with $|a|\ge 2^n|\Jx\Jy|^{n/4}|h|^{n/2}(n/2)!$.
This proves \rlb{bn>1d}.~\qedm

\subsection{Signature of quantum chaos in the imaginary-time evolution of operators}
\label{s:ITE}
As in Theorem~\ref{T:OG}, we study $S=1/2$ quantum spin system on $\Zd$ with the (formal) Hamiltonian $\hH$ of the form \rlb{HZd} corresponding to \rlb{H}.
For a positive integer $L$, let $\hH_L$ be the same Hamiltonian restricted onto the finite subset $\{-L,-L+1,\ldots,L\}^d\subset\Zd$.
Since $\hH_L$ is a finite-dimensional matrix, the corresponding time evolution of a local operator $\hB_0$
\eq
\hB_L(t)=e^{\ri\hH_L t}\,\hB_0\,e^{-\ri\hH_L t},
\lb{BLt}
\en
is well-defined for any $t\in\bbC$.
It is standard that, for any $t\in\bbR$, the infinite volume limit 
\eq
\hB(t)=\lim_{L\up\infty}\hB_L(t),
\lb{Btrig}
\en
is well-defined; it belongs to the corresponding C$^*$-algebra, and the convergence is with respect to the operator norm.

In a general one-dimensional quantum spin system with translation-invariant short-range interactions, Araki proved that the same limit \rlb{Btrig} is well-defined for any $\beta\in\bbR$  \cite{Araki1969}.
Interestingly, this is no longer the case in quantum spin systems in two or higher dimensions, as was demonstrated rigorously by Bouch \cite{Bouch2015} for a particular two-dimensional quantum spin system.
According to  Avdoshkin and Dymarsky, the ill-definedness of the limit \rlb{Btrig} for $t=\ri\beta$ with $\beta\in\bbR$ with sufficiently large $|\beta|$ can be regarded as another signature of quantum chaos \cite{AvdoshkinDymarsky2020}.

The following theorem extends Bouch's theorem to the standard and general class of models that we study in the present paper.\footnote{By using the technique developed by Cao \cite{Cao2021}, one can prove the same theorem for the quantum Ising model in two or higher dimensions.}

\begin{theorem}[Singularity in complex-time evolution]
We treat the same models as in Theorem~\ref{T:OG} and set $\hB_0=\hX_o$.
For $\beta\in\bbR$ such that $|\beta|>2/\alpha$, $\snorm{\hB_L(\ri\beta)}_{\rm op}$ diverges as $L\up\infty$.
Here $\snorm{\cdot}_{\rm op}$ denotes the operator norm.
\end{theorem}
\noindent
{\em Proof:}\/
Noting that $\sbkt{\hB_0,\hB_0}=1$, we have
\eq
\snorm{\hB_L(\ri\beta)}_{\rm op}\ge\sqrt{\sbkt{\hB_L(\ri\beta),\hB_L(\ri\beta)}}\ge\sbkt{\hB_0,\hB_L(\ri\beta)}.
\en
Since $\hH_L$ is a finite-dimensional matrix, the expansion corresponding to \rlb{fB(t)} converges absolutely for any $t\in\bbC$.
By setting $t=\ri\beta$, we have
\eq
\hB_L(\ri\beta)=e^{-\beta \hH_L}\,\hB_0\,e^{\beta\hH}=\sum_{n=0}^\infty\frac{(\ri\beta)^n}{n!}\,\hB_{L,n},
\lb{BLexp}
\en
where $\hB_{L,n}$ is defined by \rlb{BB} with $\hH$ replaced by $\hH_L$.
We thus have
\eq
\snorm{\hB_L(\ri\beta)}_{\rm op}\ge
\sum_{n=0}^\infty\frac{(\ri\beta)^n}{n!}\sbkt{\hB_0,\hB_{L,n}}.
\en
It can be easily shown\footnote{
One makes use of the fact that the Liouvillian $\delta_L$ defined by $\delta_L(\hA)=\ri[\hH_L,\hA]$ is anti-self-adjoint (with respect to the present inner product).
} that $\sbkt{\hB_0,\hB_{L,n}}=0$ for odd $n$ and $(-1)^{n/2}\sbkt{\hB_0,\hB_{L,n}}\ge0$ for even $n$.
Furthermore, since the Hamiltonian is short ranged, for any $L$ there exists $n_0(L)$ such that $\hB_{L,n}=\hB_n$ for any $n\le n_0(L)$.
We can assume $n_0(L)\up\infty$ as $L\up\infty$.
Then, by defining the moment by $\mu_{2m}=(-1)^m\sbkt{\hB_0,\hB_{2m}}$, we have the lower bound
\eq
\snorm{\hB_L(\ri\beta)}_{\rm op}\ge\sum_{m=0}^{\lfloor n_0(L)/2\rfloor}\frac{\beta^{2m}}{(2m)!}\mu_{2m}.
\en
It holds in general that $\mu_{2m}\ge(\prod_{j=1}^mb_j)^2$.  See, e.g., Appendix A2 of \cite{ParkerCaoAvdoshkinScaffidiAltman2019}.
We then find from \rlb{bn>Cn} that
\eq
\snorm{\hB_L(\ri\beta)}_{\rm op}\ge\sum_{m\in G\cap[0,n_0(L)/2]}(\alpha\beta)^{2m}\frac{(m!)^2}{(2m)!},
\en
where the sum diversges as $n_0(L)\up\infty$ if $\alpha\beta/2>1$.~\qedm

%
%
%
%

 \bigskip
\noindent{\small
{\em Acknowledgement} || 
We thank Yuuya Chiba for informing us of his results prior to publication, and Xiangyu Cao, Akihiro Hokkyo, Hosho Katsura, Chihiro Matsui, Pratik Nandy, HaRu Park, Daniel Ueltschi, and Mizuki Yamaguchi for valuable discussions and comments on the manuscript.
H.T. thanks Mahiro Futami for useful discussions in the early stage of the work.
N.S. was supported by JST ERATO Grant Number JPMJER2302 and H.T. by JSPS Grants-in-Aid for Scientific Research Nos. 22K03474 and 25K07171.}


\end{document}